\documentclass[aps,pra,reprint,amsmath,amssymb,superscriptaddress,onecolumn,longbibliography,nofootinbib,notitlepage]{revtex4-2}
\usepackage{mathtools}
\usepackage{siunitx}
\usepackage[hidelinks]{hyperref}
\usepackage{svg}
\svgpath{{Figures/}}
\usepackage{caption}
\usepackage{xfrac}
\usepackage{float}
\usepackage{subcaption}
\usepackage[braket, qm]{qcircuit}
\usepackage{bbm}
\usepackage{lipsum} 

\usepackage{soul}
\usepackage[normalem]{ulem}

\usepackage{xcolor}

\newcommand{\correspondingauthors}{%
To whom correspondence should be addressed:  ps2228@cornell.edu, mo522@cornell.edu, and pmcmahon@cornell.edu.}

\usepackage{ragged2e}

\makeatletter
\long\def\@makecaption#1#2{%
  \par
  \vskip\abovecaptionskip
  \small
  \justifying
  \noindent #1. #2\par
  \vskip\belowcaptionskip
}
\makeatother

\begin{document}
\title{Quantum sensors that compute:\\ quantum computational magnetic-field sensing using a superconducting qubit}

\author{Purnendu~Sen}
\thanks{These authors contributed equally.}
\affiliation{School of Applied and Engineering Physics, Cornell University, Ithaca, NY 14853, USA}

\author{Mathieu~Ouellet}
\thanks{These authors contributed equally.}
\affiliation{School of Applied and Engineering Physics, Cornell University, Ithaca, NY 14853, USA}

\author{Saeed~A.~Khan}

\author{Wayne~Wang}
\affiliation{School of Applied and Engineering Physics, Cornell University, Ithaca, NY 14853, USA}

\author{Sridhar~Prabhu}
\affiliation{School of Applied and Engineering Physics, Cornell University, Ithaca, NY 14853, USA}
\affiliation{Department of Physics, Cornell University, Ithaca, NY 14853, USA}

\author{Alen~Senanian}
\altaffiliation{Present address: Diraq, Chicago, IL 60649, USA}
\affiliation{School of Applied and Engineering Physics, Cornell University, Ithaca, NY 14853, USA}
\affiliation{Department of Physics, Cornell University, Ithaca, NY 14853, USA}

\author{William~P.~Banner}
\affiliation{Department of Electrical Engineering and Computer Science, Massachusetts Institute of Technology, Cambridge, MA 02139, USA}

\author{William~D.~Oliver}
\affiliation{Department of Electrical Engineering and Computer Science, Massachusetts Institute of Technology, Cambridge, MA 02139, USA}
\affiliation{Research Laboratory of Electronics, Massachusetts Institute of Technology, Cambridge, MA 02139, USA}
\affiliation{Department of Physics, Massachusetts Institute of Technology, Cambridge, MA 02139, USA}

\author{Peter~L.~McMahon}
\thanks{\correspondingauthors}
\affiliation{School of Applied and Engineering Physics, Cornell University, Ithaca, NY 14853, USA}

\begin{abstract}

A measurement of a single-qubit quantum sensor reveals at most 1 bit of information about the signal that was sensed. To perform a classification task on the signal, the conventional approach is to repeat a sensing protocol many times, averaging the measurement results to obtain a high-precision estimate of the sensed signal, and then to apply classical postprocessing.

Quantum computational sensing (QCS) is an alternative approach that breaks with the paradigm of first obtaining a classical estimate of the signal and then computing a function of the estimated signal in postprocessing. QCS instead combines quantum sensing with quantum computing to concentrate information about the signal relevant to the task into the solitary bit revealed by each measurement.
QCS trades off generality for greater efficiency, allowing a task to be completed with less sensing time than would have been required in the conventional approach.

Here, we report on the experimental demonstration of QCS where sensing and computing were both performed by the same single superconducting transmon qubit. 
We consider various binary classification tasks based on static and oscillating magnetic fields induced by current through a flux line. 
The fields were sensed through a double-junction superconducting quantum interference device (SQUID) loop that was part of the qubit. 
We used a protocol based on quantum signal processing to preprocess the sensed signals in the quantum domain prior to measurement. 
For tasks on static magnetic fields, our protocol outperformed the conventional baseline of Ramsey-based phase estimation by as much as 15 percentage points. 
For tasks on oscillating magnetic fields, we classified signal amplitude and frequency with up to 20 and 15 percentage points higher accuracy, respectively, compared to the conventional baseline of optimized dynamical-decoupling protocols. 
Our results illustrate how quantum computing can enhance quantum sensing even with a minimally sized quantum system subject to the practical limitations of decoherence and error-prone operations.

\end{abstract}

\maketitle

\section{Introduction}
\label{sec:intro}
\begin{figure*}
    \centering
    \includegraphics[width=\linewidth]{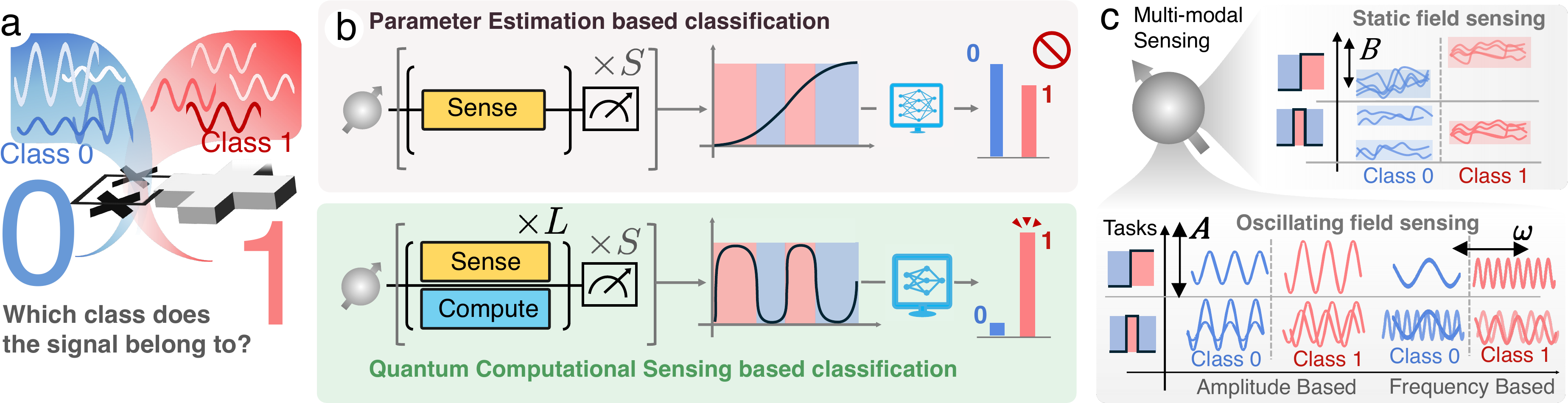}
    \caption{
    \textbf{
    Conventional quantum sensing versus quantum computational sensing. a,} 
    Example of a one-dimensional magnetic field sensing task where the goal is to extract relevant features from a time-dependent signal $u(t)$. 
    A magnetic field sensor measures the field perturbed by a nearby unknown object, and the task is to determine whether the signal originates from class 0 or class 1. 
    \textbf{b,} (Top) Conventionally, a quantum sensor measures the magnetic field as a time-series. The protocol is repeated for \textit{S} shots to get reliable estimates of the expectation value.
    The measured signal is then processed classically, for example with a neural network, to extract features that allow identifying the object producing the field. This corresponds to computing the function that maps the measured signal to the correct object label classically. 
    (Bottom) A quantum computational sensor combines sensing and computation to perform Quantum Computational Sensing (QCS) within the quantum system to directly produce a function at the output, rather than solely estimates of the raw magnetic field. The sensing operations are interleaved with control unitaries for \textit{L} layers, and the whole protocol is repeated for \textit{S} shots.
    In this case, the sensor outputs the class label of the object, and minimal classical postprocessing is required, allowing a more efficient use of quantum resources.  
    \textbf{c,}    
    Examples of classification tasks: DC signal detection, AC amplitude classification, and AC frequency classification with two tasks of increasing complexity. 
    As complexity increases, the signal parameter space is divided into progressively smaller regions that correspond to different binary classes, making the classification task more demanding. 
    }
    \label{fig:fig1}
\end{figure*}

Quantum sensing uses quantum systems to estimate the value of sensed physical quantities (which are often called sensed parameters, and which we will refer to as the sensed signal) \cite{marciniak_optimal_2022, kolosvetov_quantum_2026, danilin_quantum_2024}. 
Quantum computational sensing (QCS)~\cite{eldredge_optimal_2018, zhuang_physical-layer_2019, banchi_quantum-enhanced_2020, quantum_computational_imaging, debry_experimental_2023,  sinanan-singh_single-shot_2024, liao_quantum-enhanced_2024, khan_quantum_2025, khan_quantum_2025-1} combines quantum sensing and quantum computing to more efficiently learn task-specific information about one or more sensed signals. 
In conventional quantum sensing, a sensor is prepared in an initial state and allowed to evolve under an interaction that depends on the sensed signal, denoted by $u(t)$. 
This evolution encodes information about $u(t)$ in the final sensor state.
A measurement of this state produces outcomes from which the sensed quantities are estimated~\cite{degen2017quantum}. 
For a given measurement, not all information encoded in the quantum state is necessarily accessible, and finite sampling limits the precision with which the sensed quantities can be estimated \cite{hu_tackling_2023, meyer_quantum_2025}.
When the goal is to infer a task-specific property $F^\star(u)$ of the signal, such as a class label, QCS provides an alternative approach (Fig.~\ref{fig:fig1}).
Instead of first constructing an estimate of the sensed signal and then postprocessing the result classically, QCS incorporates task-specific computation into the sensing protocol itself, mapping information about $F^\star (u)$ onto the outcome probabilities of the final measurement.
This task-aware approach can enable QCS to outperform sensing protocols based on parameter estimation, even in small-scale quantum systems. A quantum computational sensing advantage (QCSA)~\cite{khan_quantum_2025} can be realized using only a single qubit.

In this work, we experimentally demonstrate QCS using a single superconducting transmon and show classification advantages over optimized conventional sensing protocols.
We consider binary classification tasks for static and oscillating magnetic fields defined by a family of labeling functions ${F^\star(u)}$ of increasing complexity, with the signal parameter space divided into progressively smaller regions.
Rather than estimating the signal and classically evaluating ${F^\star(u)}$ after measurement, we perform quantum signal processing (QSP) ~\cite{low_optimal_hamiltonian_sim,low_composite_quantum_gates, martyn2021grand, liu2025toward, martyn2025parallel}, a sequence of coherent control operations interleaved with sensing unitaries for \textit{L} layers, prior to measurement to encode task-relevant information more directly into the measurement outcomes. The protocol is repeated for $S$ shots, whose average is thresholded to read out the class encoded in the measurement probability.
QSP-based QCS protocols have been proposed~\cite{allen_quantum_2025,khan_quantum_2025-1,sinanan-singh_single-shot_2024,liao_quantum-enhanced_2024,prabhu_khan_QCDS_2026} for various tasks in both qubit-only and hybrid qumode-qubit settings.

Our system comprises a frequency-tunable transmon whose g-e transition frequency depends on the applied magnetic flux. 
The applied flux induces a $\sigma_z$ rotation of the quantum state via an effective detuning between the gate
pulses and the qubit frequency~\cite{krantz_quantum_2019, flux_tunable_transmon} (see Appendix~Fig.~\ref{app_fig:calibration} and Appendix~\ref{app_sec:experimental_setup}).
The QCS protocols experimentally implemented here are adapted from the QSP protocols theoretically proposed for binary classification in Ref.~\cite{khan_quantum_2025-1}.
As described in Appendix~\ref{app_sec:DC_QCS}, we extended these protocols to continuously applied signals by accounting for the signal-induced detuning during control operations and optimizing them in the presence of stochastic phase noise.
We train the sequence of unitary operations using supervised learning to maximize correlation between finitely sampled measurement outcomes and the class labels defined by the target function $ F^\star(u) $.
We perform binary classification of static magnetic fields based on the field strength. 
For oscillating fields, we consider two classification tasks, where $u(t)$ encodes either the signal amplitude or its frequency \cite{kristen_amplitude_2020}, and demonstrate an advantage in both cases.
We benchmark the QCS protocol against a Ramsey-based conventional sensing baseline followed by classical post-processing
for static fields and against Dynamical-Decoupling (DD)~\cite{taylor_high-sensitivity_2008,de_lange_single-spin_2011,naydenov_dynamical_2011,hall_ultrasensitive_2010,maze_nanoscale_2008} protocols for oscillating fields, and observe that QCS enables higher classification accuracies across a family of tasks of increasing complexity.
\section{Experimental Protocols and Results}
\begin{figure*}[htb]
    \centering
    \includegraphics[width=\linewidth]{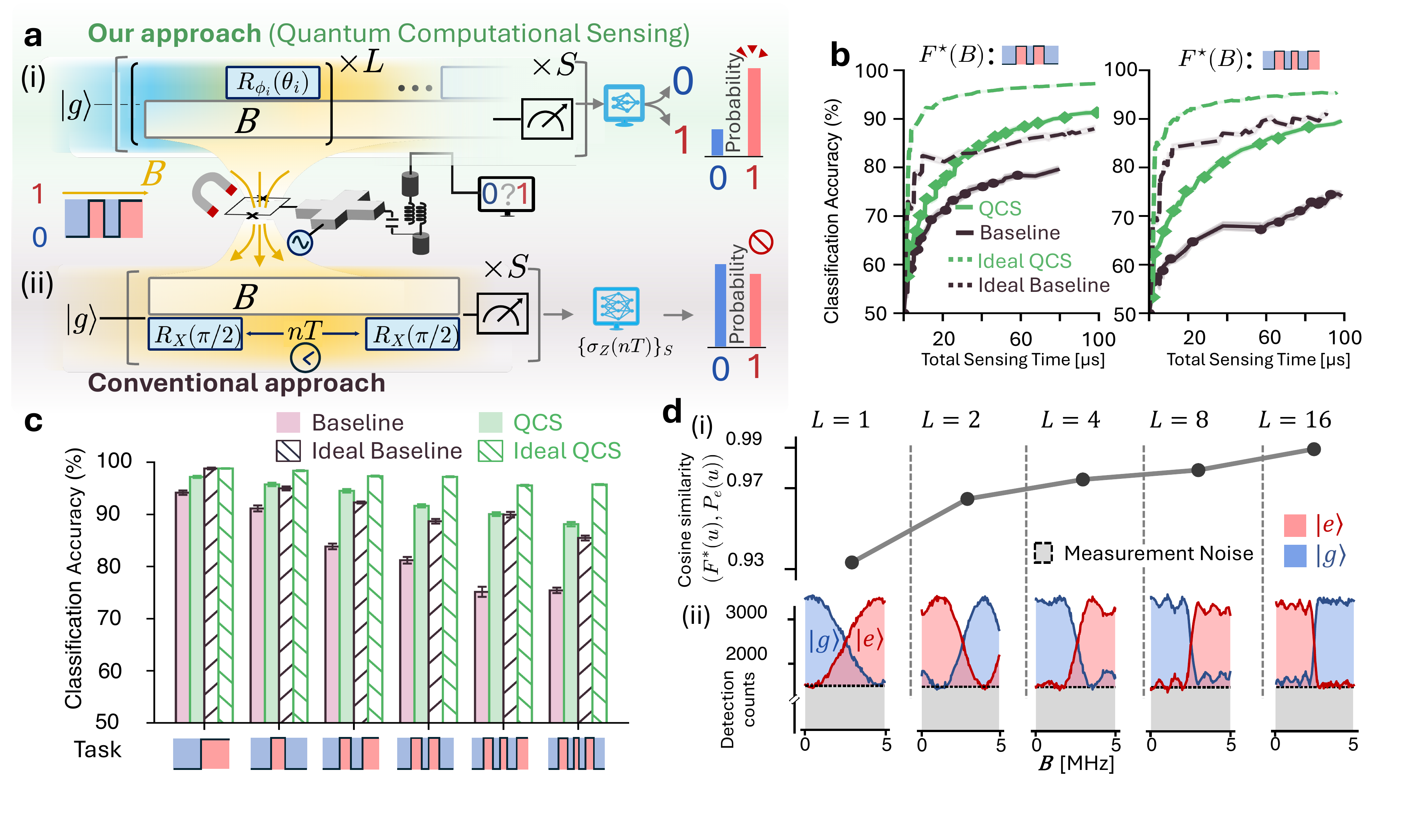}
    \caption{
    \textbf{
    Static magnetic-field sensing: Amplitude classification. a,} Comparison between a conventional sensing baseline and the quantum computational sensing (QCS) protocol. 
    (i)  The QCS protocol comprises an alternating sequence of controlled rotations and sensing intervals. 
    Each sensing period allows the signal to accumulate phase, while the rotations transform the quantum state according to a programmed sequence. 
    After $L$ layers of rotation-wait operations, the resulting measurement implements a nonlinear function of the sensed signal. The protocol is repeated for $S$ shots.
    (ii) 
    The conventional sensing protocol comprises a series of Ramsey measurements, where the parameter is encoded as a phase. The protocol is repeated for a series of interrogation times $T/2$, $T$, ..., where each measurement progressively refines the estimate of the magnitude of the underlying DC field. Each Ramsey measurement was repeated over multiple shots ($S$), with the shot allocation optimized to maximize classification accuracy.
    \textbf{b,}
    Example of two classification tasks of differing complexity. 
    The DC signals spanned qubit frequency detunings from 0 to 5 MHz, with sensing times up to 100 \si{\micro s}.
    By programming the sensing sequence, the QCS protocol encodes more task-relevant information into the final measurement basis before the final destructive measurement of the qubit.
    Experimental results are compared with ideal, shot-noise-limited QCS and the conventional baseline under the same sensing-time constraints.
    \textbf{c,}
    Classification accuracy as a function of task complexity, showing improvement across a broad range of regimes. 
    The total sensing time is fixed to 100 \si{\micro s}.
    QCS consistently outperforms the experimental baseline and, except for the simplest task, also surpasses ideal, shot-noise-limited performance of the baseline by up to 3.0 percentage points.
    In \textbf{b} and \textbf{c}, shaded regions and error bars, respectively, indicate $\pm1$ standard deviation across 20 computational resamplings of the measurement outcomes for the same trained protocol.
    \textbf{d,}
    (i) Cosine similarity between $P_e(B)$ and the class labels for different numbers of layers, showing that the similarity increases as the protocol becomes deeper, indicating that the sensor computes a function closer to the target function.
    (ii) Distribution of the measured $\ket{g}$ (blue) and $\ket{e}$ (red) outcomes, with the single-shot measurement uncertainty shown in grey. As the number of layers increases, the distributions become progressively sharper at the transition boundary between classes.
    }
    \label{fig:fig2}
\end{figure*}

\subsection{Static magnetic field sensing}
\label{sec:stat_mag_field}
We considered a binary classification task to demonstrate the advantage of computation prior to measurement. 
Each task is defined by a one-dimensional labeling function $F^\star(u)$ over the magnetic-field range, which assigns one of two possible class labels to each field value.
For Task $n$, with $n=1,\ldots,6$, we normalize the signal parameter as
$x=(u-u_{\min})/(u_{\max}-u_{\min})$. The class boundaries are defined by
\begin{equation*}
    c_{n,k}
    =
    \frac{k}{n+1}
    +
    \frac{1}{20}
    \sin\left(\frac{2\pi k}{n+1}\right),
    \qquad k=0,\ldots,n+1.
\end{equation*}
The labeling function is then given by
\begin{equation*}
    F_n^\ast(u)=k \bmod 2,
    \qquad
    c_{n,k}\leq x<c_{n,k+1},
\end{equation*}
for $k=0,\ldots,n$. Throughout this work, $u$ was sampled from a balanced distribution obtained by uniformly sampling over the signal range and rejecting samples as needed to retain equal representation of both classes.
For all QCS results, classical postprocessing consisted of averaging the binary outcomes and applying a decision rule that, in nearly all cases, reduced to a single threshold implemented using a random-forest classifier.
All protocols considered in this work were non-adaptive: the control parameters and shot allocations were fixed before data acquisition and did not depend on outcomes from previous shots.

The control unitaries are trained in silico on a digital twin of the system (see Appendix~\ref{app_sec:DC_QCS}) to map the final qubit state onto a measurement-basis eigenstate determined by the class of the magnetic field.
A frequency-tunable double-junction transmon evolves under an incoming magnetic field $u(t)$ interleaved with a series of control unitaries as depicted in Fig.~\ref{fig:fig2}a. 
Throughout this work, the magnetic-field signal has the general form
\begin{equation*}
    u(t)=A\cos(\omega t+\phi)+B,
\end{equation*}
and an additional static bias sets the qubit at an operating point where $u(t)$ appreciably affects its frequency.
For the time-independent case, we set $A=0$, such that $u(t)=B$ is constant and serves as the task parameter.

We define the total sensing time $T_s$ as the cumulative protocol duration over all measurement shots. 
For each shot, this includes the control pulses, the intervening free-evolution intervals, and the readout, because the signal remains continuously applied throughout the protocol, including during readout. 
Consequently, readout is performed in the presence of the unknown signal using fixed readout settings. 

We implemented a sequence of Ramsey measurements with interrogation times $T/2$, $T$, $\ldots$, which progressively refine the estimate of the underlying DC magnetic field, providing a baseline for comparison with our computational-sensing protocol (see Fig.~\ref{fig:fig2}a and Appendix~\ref{app_sec:DC_baseline}). 
The measurement records were then postprocessed with a classifier to evaluate classification accuracy for different shot allocations.
Conventional phase-estimation protocols commonly use exponentially increasing interrogation times to resolve successive binary digits of the signal~\cite{dobvsivcek2007arbitrary}. 
This produces regular binary subdivisions of the signal range that are generally not aligned with the class boundaries considered here. 
We therefore implemented a multitime Ramsey baseline using linearly spaced interrogation times that better resolve the structure of these boundaries (see Fig.~\ref{fig:fig2}a and Appendix~\ref{app_sec:DC_baseline}). 
The measurement records were jointly postprocessed with a classifier, and the shot allocation across interrogation times was optimized separately for each task.

We assessed the QCS protocol across six labeling rules $F^\star(u)$ of increasing complexity over the same signal range. 
For each task, we swept the total sensing-time budget and used the validation set to select the configuration with the highest accuracy among those satisfying that budget. 
For QCS, this optimization included the pulse sequence, number of layers $L$, and number of shots $S$; the corresponding protocol parameters and shot allocation were optimized similarly for the conventional baseline.
The resulting Pareto fronts therefore show the highest classification accuracy achievable by each approach for a fixed duration of access to the signal. Selected configurations were then evaluated on the independent test set.

Across all binary classification tasks considered, the QCS protocol outperformed the conventional baseline.
For illustration, Fig.~\ref{fig:fig2}b shows the test-set Pareto fronts for two representative tasks over sensing times from 10 to 100 \si{\micro\second}, together with those of ideal, shot-noise-limited performance of the baseline and QCS under the same time constraints.
Across the six tasks, the Pareto fronts span low-budget protocols with $L=2$--$14$ and $S=4$--$7$ to higher-budget protocols with $L=4$--$16$ and $S=29$--$46$, demonstrating an efficient allocation of resources between coherent processing depth and repeated measurements.
The corresponding results for the remaining tasks, together with a comparison with a conventional Ramsey protocol, are provided in Appendix Fig.~\ref{app_fig:dc_all_task}.
We summarize the classification accuracy at a fixed sensing time of 100 \si{\micro\second} in Fig.~\ref{fig:fig2}c.
We observed that the QCS protocol maintained a consistent accuracy advantage of 3.0 to 14.9 percentage points over the conventional sensing baseline followed by classical postprocessing, with greater QCSA observed for more complex tasks.
Our experimental implementation of QCS also achieved performance comparable to ideal shot-noise-limited implementation of the conventional baseline, with differences of up to 3.0 percentage points.
Importantly, the QCS protocol also outperformed the baseline under ideal, shot-noise-limited conditions for both protocols, across all but the simplest task, with gains ranging from 3.4 to 10.3 percentage points.

The improved performance is consistent with QCS avoiding the intermediate estimation of a continuous parameter, allowing us to directly encode $\langle F^\star(u) \rangle$ into the measurement outcome, rather than first estimating $u$. 
In conventional sensing, information about the magnetic field is first inferred from projective measurements and only then assigned to a class, which can discard task-relevant information \cite{giovannetti_quantum-enhanced_2004, giovannetti_advances_2011}.
In contrast, QCS directly maps the measured state towards an eigenstate of the readout basis associated with the target class, so that the classification-relevant information is encoded directly in the measurement outcome, limiting the loss of information. 
For fairness, the baseline protocols were also optimized over accessible control parameters to perform as much of the classification task as possible during sensing, rather than acting as pure parameter estimators followed by external postprocessing.
In particular, the conventional baseline was optimized over shot allocations across the available wait times, with task-specific allocations chosen to maximize the information extracted during sensing (see Appendix~Fig.~\ref{app_fig:dc_baseline_training}).

To effectively map the measured state to a class label in QCS protocols, we require sufficient circuit depth and expressivity~\cite{schuld_circuit-centric_2020}.
We quantified the scale-invariant alignment between the excitation probability of the qubit ($P_e$) and the target function $F^\star(u)$ using their cosine similarity, 
which is the normalized inner product of the realized function $P_e$ and the target function over a range of samples $\{u_k\}$.
Figure~\ref{fig:fig2}d shows the cosine similarity between the measured $P_e$ and target $F^\star(u) $.
As the number of interleaved control layers increases, the measured response more closely follows the target class boundary. 
For the given range of tasks, we empirically observed that the number of layers required to reach a given error increases approximately linearly with task number up to approximately 18, beyond which gains become marginal (see Appendix~Fig.~\ref{app_fig:DC_training}).

\subsection{Oscillating magnetic field sensing}
\subsubsection{Amplitude classification}

\begin{figure}[htb]
    \centering
    \includegraphics[width=\linewidth]{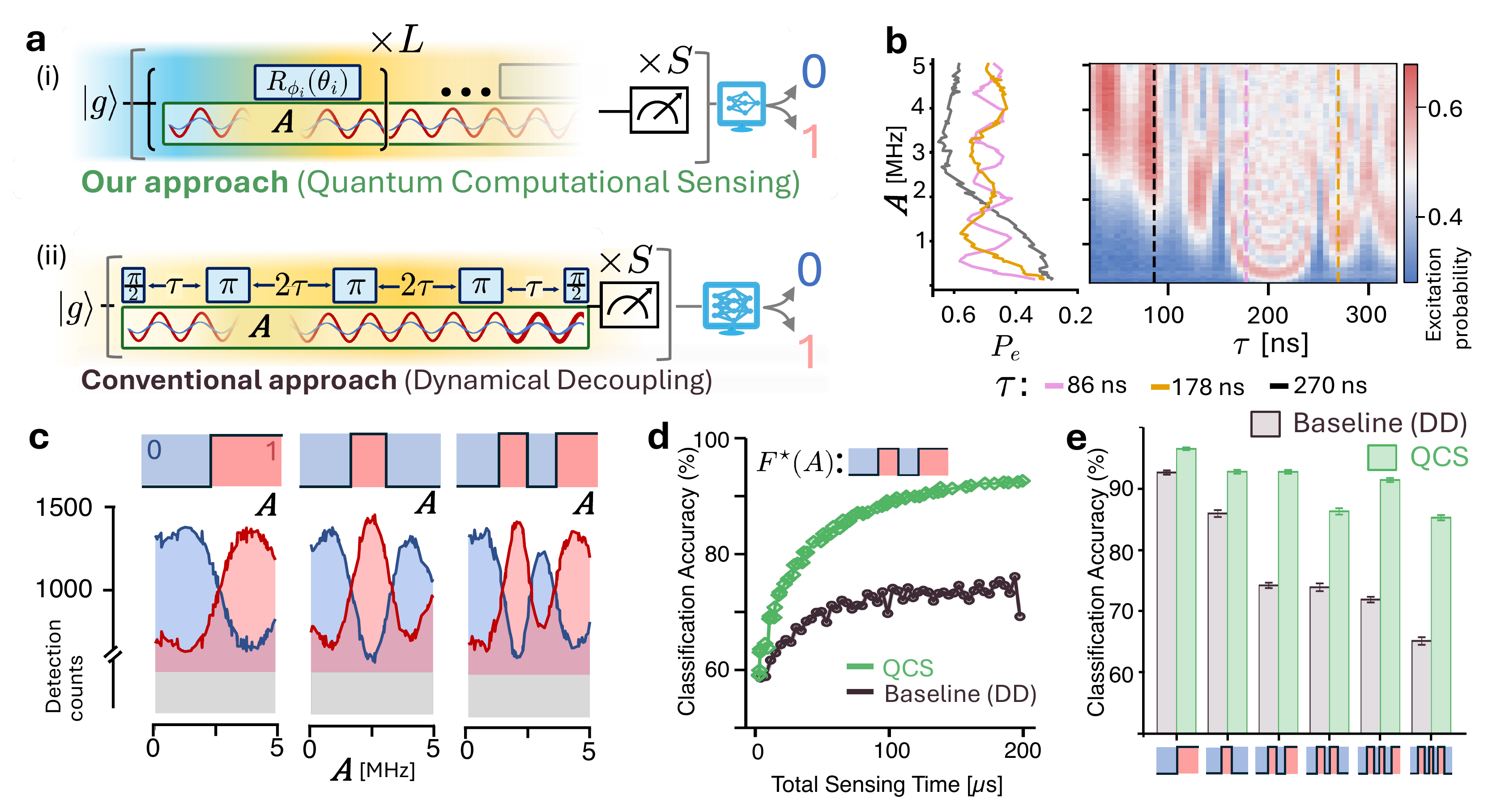}
    \caption{
    \textbf{Oscillating magnetic-field sensing: Amplitude classification. a,} Comparison between a traditional dynamical decoupling protocol (DD) and the quantum computational sensing (QCS) protocol for sensing the amplitude of an oscillating magnetic field.
    (i)  The QCS protocol comprises a sequence of programmed rotations alternating with sensing intervals. During each interval the AC signal accumulates phase, with a random initial phase at the start of the protocol. After $L$ layers, the measurement implements a function of the signal amplitude.
    (ii) The DD protocol comprises a sequence of $N$ $\pi$-pulses separated by waiting periods 
    $\tau$ that modulate the qubit evolution. The accumulated phase encodes properties of the AC signal, such as its amplitude.
    \textbf{b,} $P_e$ output of the DD sequence for $N=10$, for signal amplitudes from $0$ to $5$ MHz and waiting times $\tau$ from 0 to 325 ns, showing the nontrivial function of the signal that can be computed.
    (left) Cross sections of $P_e $ for three different waiting times, illustrating the functions that can be computed.
    (right) Heatmap of the dynamical decoupling sequence showing the probability of exciting the qubit ($P_e$) as a function of the incoming signal strength and wait times.  
    \textbf{c,}
    Distribution of the measured $\ket{g}$ (blue) and $\ket{e}$ (red) outcomes, with measurement noise shown in grey, for three different tasks.
    \textbf{d,}
    Example of a classification task shown in the figure: AC signals with amplitudes between 0 and 5 MHz are sensed over a window from 0 to 200 \si{\micro\second}, showing a large improvement over the best DD protocol for this task.
    \textbf{e,}
    Classification accuracy as a function of task complexity, showing improvement across a broad range of regimes. 
    The total sensing time is fixed to 200 $\si{\micro\second}$. 
    Error bars show $\pm1$ standard deviation across 20 computational resamplings of the measurement outcomes for the same trained protocol, each using the corresponding number of measurement shots.
    }
    \label{fig:fig3}
\end{figure}
Next, we considered the classification of oscillating signals in the phase-incoherent regime \cite{grochowski_dist_phase_insens_sensing, isogawa_entanglement-assisted_2026}, where each experimental sequence sampled a different random phase.
Here, $B=0$, $\omega/2\pi$ was fixed at $2~\mathrm{MHz}$, and $A$ served as the task parameter. 
Each experimental sequence sampled a different random phase $\phi$, and the double-junction transmon performed amplitude classification using the same family of labeling functions $F^\star(u)$ introduced in Section~\ref{sec:stat_mag_field} (Fig.~\ref{fig:fig3}a).
The class label is determined by the amplitude of the sinusoid rather than by the instantaneous signal value during sensing.
The QCS protocol follows the same structure as in the static case but depends on the instantaneous phase $\phi$, which varies randomly between shots (refer to Appendix~\ref{app_sec:AC_QCS} for details about the protocol and training of the required control unitaries $U_{u,\phi}(t)$).
We benchmarked the protocol against a Dynamical Decoupling (DD) protocol optimized for each classification task \cite{taylor_high-sensitivity_2008,de_lange_single-spin_2011,naydenov_dynamical_2011,hall_ultrasensitive_2010,maze_nanoscale_2008}.
We simulated DD sequences with varying numbers of $\pi$-pulses and interpulse delays up to 500 ns, and selected the six best-performing sequences for experimental tests (see Appendix~\ref{app_sec:AC_DD} for details about the training and selection of optimal DD sequences).
For each total sensing time, we selected the optimal experimental protocol from the validation-set Pareto front, with representative examples shown in Fig.~\ref{fig:fig3}b.

QCS outperformed the baseline for signal classification across the full experimental range. 
Figure~\ref{fig:fig3}c shows that QCS suppresses the sensitivity to random phase and, as in the time-independent case, successfully performs the classification task (see Appendix Fig.~\ref{app_fig:supp_readout}a). 
Figure~\ref{fig:fig3}d shows the test-set Pareto frontier of both protocols for this task, indicating that QCS outperforms DD over the entire range of sensing time from 0 to 200 \si{\micro\second}. 
The Pareto-optimal strategy evolves from shallow, single-shot sensing ($L=5$--$7$, $S=1$) to deeper protocols supported by repeated measurements ($L=8$--$22$, $S=42$--$68$) as the sensing-time budget increases.
Across all task complexities, the observed improvement ranges from 3.8 to 20.1 percentage points, with greater advantage observed for more complex tasks, and classification accuracies spanning from 85.3\% to 96.5\%.
The optimized protocols, task complexities, and corresponding training curves are shown in Appendix Fig.~\ref{app_fig:supp_ac_all}.

The observed DD performance is traceable to the restricted computational structure available within these protocols.
The DD protocols provide limited input-output mappings that can be implemented during sensing, allowing for either Ramsey-like or related oscillatory features in the measured response as shown in Fig.~\ref{fig:fig3}b (see Appendix Fig.~\ref{app_fig:supp_readout}c). 
However, because dynamical-decoupling protocols are not specialized to classify time-varying signals separated by nonlinear decision boundaries they fail to reproduce the more complicated $F^\star(u)$ functions, which contain fine, non-periodic features across the signal space.
This explains why DD remains competitive on simpler tasks, where only coarse distinctions are required. 
Importantly, the observed advantage is not driven by readout alone, but also by the greater range of signal-response functions of the protocol. Our QCS protocol outperformed DD across various readout fidelities, spanning from $50\%$ to perfect readouts 
(Appendix Fig.~\ref{app_fig:supp_readout}b).

\begin{figure}[htb]
    \centering
    \includegraphics[width=\linewidth]{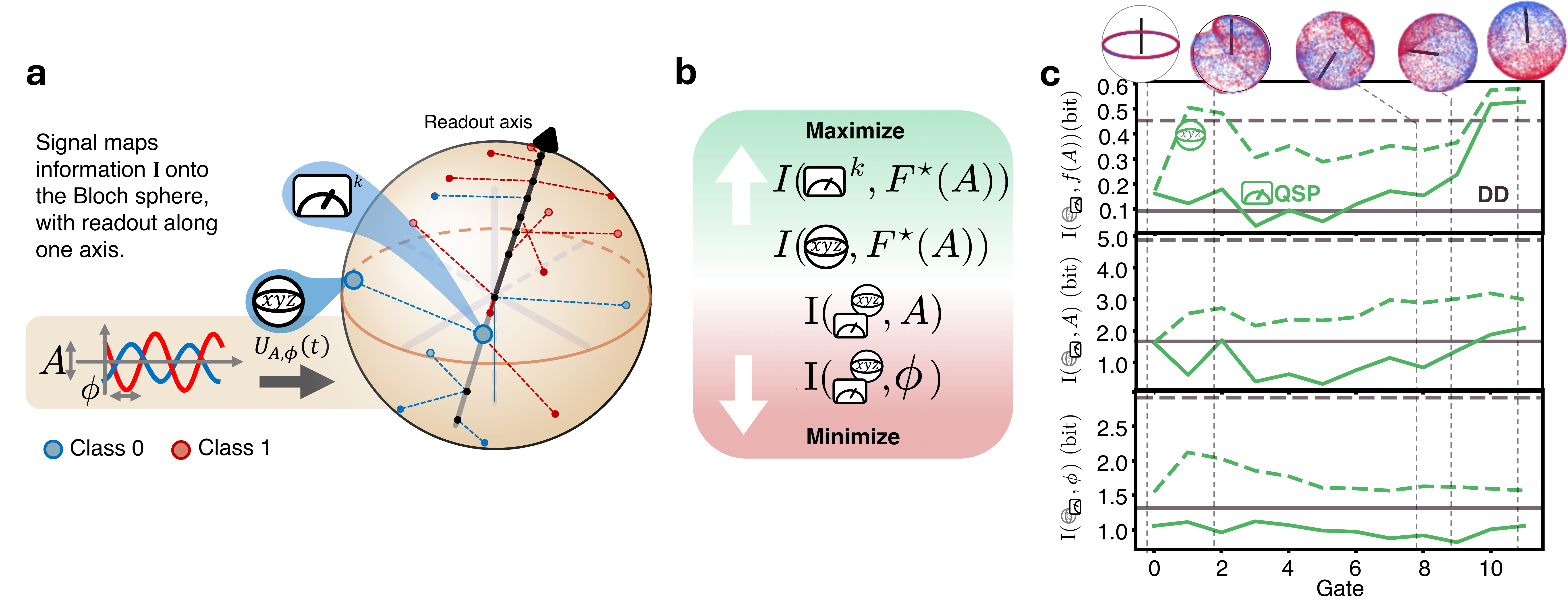}
    \caption{ \textbf{Encoding task-relevant information in measurement outcome.}
    \textbf{a,} An input signal parameterized by amplitude $A$ and phase $\phi$ is encoded into the qubit evolution through a sequence of controlled unitary operations $U_{A,\phi}(t)$, generating trajectories on the Bloch sphere. Different signal classes (blue and red) evolve into distinguishable regions when projected onto the measurement axis. \textbf{b,}  The QSP-based sensing protocol is trained to maximize the mutual information between the projective measurement outcomes and the target function $F^\star(A)$, while suppressing sensitivity to phase.  \textbf{c,} Mutual information evolution across the QSP sequence for different quantities, demonstrating enhanced correlation between the measurement outcomes and the desired classification target relative to dynamical decoupling (DD) protocols.
    }
\label{fig:fig_info}
\end{figure}
The computational sensing advantage in classifying phase-incoherent oscillatory signals is associated with the increased mutual information between the class label $F^\star(u)$ and the measurement outcome (see Fig.~\ref{fig:fig_info}a).
Early in the protocol, mutual information about the task $F^\star(u)$ is already present in the full Bloch vector, but this information is not yet accessible through a single projective measurement (see Fig.~\ref{fig:fig_info}b).
The mutual information in the measurement basis remains low during the early layers, indicating that these layers primarily prepare and reshape the state before task-relevant information is concentrated onto the measurement axis (see Appendix~\ref{app_ssec:information_theory}). 
By the end of the protocol, most of the mutual information in the full Bloch vector is concentrated in the measurement basis, as shown in Fig.~\ref{fig:fig_info}b.
In contrast, the measurement basis remains only weakly informative about the amplitude $u$, despite the larger amount of amplitude information retained in the full Bloch vector. 
Information about the phase $\phi$ is also largely absent from the measurement basis and is progressively suppressed in the full state, as expected for a phase-agnostic task.

\subsubsection{Frequency classification}
\begin{figure}[htb]
    \centering
    \includegraphics[width=0.9\linewidth]{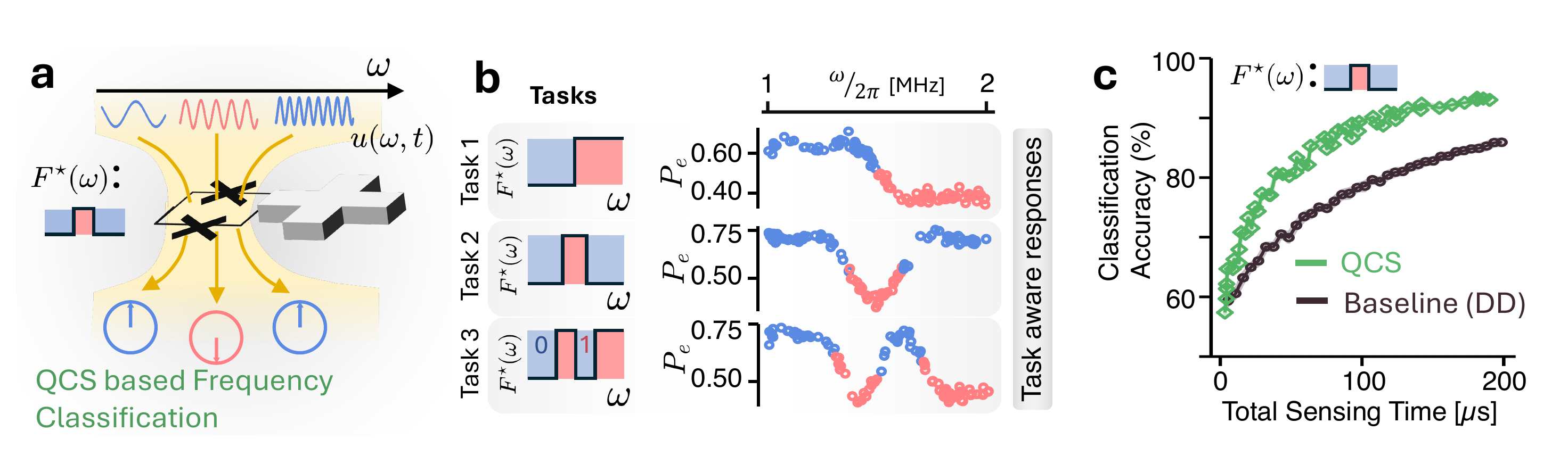}
    \caption{
    \textbf{Oscillating magnetic-field sensing: Frequency classification. a,}  Using a similar protocol as in Fig.~\ref{fig:fig3}, we now consider tasks involving classification of the frequency for signals ranging from 1 MHz to 2 MHz.
    \textbf{b,}
    Examples of  classification tasks are shown in the figure: AC signals with frequencies between 1 and 2 MHz are sensed over a window from 0 to 200 \si{\micro s}, showing an improvement over the best DD protocol for this task.
    \textbf{c,} Classification accuracy improvement obtained with the QCS protocol over an optimized DD sequence for the second task, showing that performing computation within the quantum system improves the accuracy. 
    }
    \label{fig:fig5}
\end{figure}

Next, we considered the classification of time-dependent signals based on their frequency, and showed how computational sensing performs in a regime where dynamical decoupling is particularly effective (see Appendix~\ref{app_ssec:DD_training}). 
Here, $B=0$, $A$ was fixed at $\mathrm{0.6}~\mathrm{MHz}$, $\omega$ served as the task parameter, and each experimental sequence sampled a different random phase $\phi$.
We considered sinusoidal signals with a 1 MHz bandwidth centered at 1.5 MHz, and classified their frequency using the same family of labeling functions across six task complexities.
As in amplitude classification, the QCS protocol compensated for the random phase and reproduced functions that closely follow the labeling functions as shown in Fig.~\ref{fig:fig5}b.

Figure~\ref{fig:fig5}c shows the test-set Pareto fronts for Task~2, revealing a strong QCS advantage across a range of total sensing times. 
Across all task complexities, we observed gains of up to 15.4 percentage points, while in the worst case performance decreased by 2.8 percentage points. Accuracies ranged from $81.9\%$ to $93.8\%$ as shown in Appendix Fig.~\ref{app_fig:ac_all_task}.
Across all frequency-classification tasks, for smaller sensing times, the Pareto front consists of single-shot protocols ($S=1$) with shallow depths of $L=5$--$7$. 
As the sensing-time budget increases, the optimal protocols use both greater depth and more repetitions, reaching $L=12$--$28$ and $S=32$--$43$ at the high-budget end.
While QCS underperforms DD for the first task, which uses a single threshold to divide the signal range into two classes, it outperforms DD for the more complex tasks involving multiple disjoint intervals. 
This is consistent with DD being naturally well-suited to narrow-band frequency discrimination, while QCS becomes advantageous when the classification task requires more involved processing (see Appendix Fig.~\ref{app_fig:supp_dd_exp}).

\section{Discussion}
\label{sec:discussion}

\textit{Summary of our work.} Quantum computational sensing (QCS) based on quantum signal processing (QSP) interleaves sensing operations with trained single-qubit unitary operations to compute functions of the sensed signal $u(t)$; measurement reveals information about a function of $u(t)$ rather than directly $u(t)$ itself, which allows for more efficient sensing if the ultimate goal is to estimate a function of $u(t)$.
We demonstrated the use of a QSP-based protocol with a single qubit to achieve higher accuracies in classifying static and oscillating magnetic fields compared to conventional quantum sensing methods with classical postprocessing, when given access to the signal for the same amount of time. 
We found that our QSP-based protocol was able to classify static magnetic fields with accuracies up to $14.9$ percentage points higher than a baseline based on a conventional sensing protocol, for equal total sensing durations of 100 \si{\micro\second}.
For oscillating magnetic field signals, we obtained advantages of up to $20.1$ and $15.4$ percentage points for amplitude and frequency classification tasks, respectively, at a total sensing time of 200 \si{\micro\second}.
These advantages are correlated with the enhanced mutual information between projective measurement outcomes and the underlying decision function, mitigating measurement-induced information loss.

\textit{Outlook.} For quantum computational sensing to become practically useful, it will likely require a platform that is both a state-of-the-art sensor for a given task and has sufficient coherence and controllability to allow the faithful execution of task-specific quantum protocols. Existing quantum sensing platforms, including NV centers for magnetic-field sensing for biomedical applications \cite{aslam_quantum_2023, jin_four_2026, taylor_biological_2013}, SQUID-based systems \cite{fagaly_superconducting_2006} for magnetoencephalography \cite{gross_magnetoencephalography_2019}, and atomic-vapor magnetometers for navigation \cite{quantum_assured_magnav} could each, for example, potentially be adapted to perform QCS.
While our results show a clear advantage for the synthetic tasks we considered, an important open question is for which more realistic tasks QSP-based QCS can also deliver advantages. We would like to see advantages for tasks that are more sophisticated than binary classification, and multiclass classification problems also suggest the use of multiple qubits so that a single measurement of the system can in principle contain enough information to encode the answer. 
The multiqubit setting may also be interesting to explore by allowing signals at different spatial locations to be sensed, making the function being computed have a multidimensional rather than scalar input. 
The use of multiple qubits could also enable more expressive computations, allowing for more complicated functions to be computed~\cite{khan_quantum_2025-1}, including multivariable polynomial transformations~\cite{rossi_multivariable_2022, QSVT}. 
Open questions include how to train multiqubit protocols efficiently \cite{ragone_lie_2024, QSP_phase_factor_eval} and how robust they are to realistic noise.

Because of limitations in our experimental setup, we restricted this study to non-adaptive protocols. 
An important direction for future work is to compare QSP-based QCS and conventional baselines each in an adaptive setting, where measurement outcomes from previous shots determine the protocol applied in subsequent shots.

Our results show that quantum computational sensing can be performed using merely a single qubit, which performs both the sensing and the computing. We anticipate that there may be many other examples of computational sensing tasks and single-qubit and multi-qubit systems where sensing and computing could also be efficiently combined in a similar way.

\section*{Data and code availability}
All experimental and simulation data and code used in this work are available at  \url{https://doi.org/10.5281/zenodo.21922738}.

\section*{Acknowledgements}

The authors would like to thank Vladimir Kremenetski, Saswata Roy, Mandar Sohoni, Logan Wright, Fan Wu and Ryotatsu Yanagimoto for helpful discussions and comments. The authors gratefully acknowledge MIT Lincoln Laboratory and the SQUILL Foundry for fabricating and packaging the superconducting device. This work was supported by the Air Force Office of Scientific Research under Award No. FA9550-22-1-0203, and the Army Research Office under Award No. W911NF-25-1-0261. The work at MIT was supported by the Army Research Office under Award No. W911NF-23-1-0045. M.O. acknowledges support from the Fonds de recherche du Québec Postdoctoral Research Scholarship, Grant No. 366727. The authors thank NTT Research for its financial and technical support.

\section*{Author contributions}
P.S. and M.O. designed and carried out the experiments, performed the numerical simulations, and trained the quantum sensing protocols. S.A.K. and S.P. developed the quantum-signal-processing-based protocol on which this work is based. W.W., S.P., and A.S. contributed to the experimental work. A.S. designed the superconducting device with contributions from W.P.B. and guidance from W.D.O.. P.S. and M.O. wrote the manuscript with input from all authors. P.L.M. supervised the project.

\bibliographystyle{mcmahonlab}
\bibliography{references}

\begin{thebibliography}{10}

\bibitem{marciniak_optimal_2022}
C.~D. Marciniak, T.~Feldker, I.~Pogorelov, R.~Kaubruegger, D.~V. Vasilyev, R.~van Bijnen, P.~Schindler, P.~Zoller, R.~Blatt, and T.~Monz, Optimal metrology with programmable quantum sensors.
\newblock {\em \href{https://www.nature.com/articles/s41586-022-04435-4}{Nature}} \href{https://www.nature.com/articles/s41586-022-04435-4}{{\bfseries 603}, 604--609} (2022).

\bibitem{kolosvetov_quantum_2026}
A.~Kolosvetov, N.~Gusarov, V.~Slepnev, M.~Perelshtein, V.~Sevriuk, A.~Gubaydullin, and V.~Vinokur, Quantum metrology based on superconducting qubits.
\newblock {\em \href{https://doi.org/10.1038/s41534-026-01285-0}{npj Quantum Information}} (2026).

\bibitem{danilin_quantum_2024}
S.~Danilin, N.~Nugent, and M.~Weides, Quantum sensing with tunable superconducting qubits: optimization and speed-up.
\newblock {\em \href{https://iopscience.iop.org/article/10.1088/1367-2630/ad49c5}{New Journal of Physics}} \href{https://iopscience.iop.org/article/10.1088/1367-2630/ad49c5}{{\bfseries 26}, 103029} (2024).

\bibitem{eldredge_optimal_2018}
Z.~Eldredge, M.~Foss-Feig, J.~A. Gross, S.~L. Rolston, and A.~V. Gorshkov, Optimal and secure measurement protocols for quantum sensor networks.
\newblock {\em \href{https://link.aps.org/doi/10.1103/PhysRevA.97.042337}{Physical Review A}} \href{https://link.aps.org/doi/10.1103/PhysRevA.97.042337}{{\bfseries 97}, 042337} (2018).

\bibitem{zhuang_physical-layer_2019}
Q.~Zhuang and Z.~Zhang, Physical-{Layer} {Supervised} {Learning} {Assisted} by an {Entangled} {Sensor} {Network}.
\newblock {\em \href{https://link.aps.org/doi/10.1103/PhysRevX.9.041023}{Physical Review X}} \href{https://link.aps.org/doi/10.1103/PhysRevX.9.041023}{{\bfseries 9}, 041023} (2019).

\bibitem{banchi_quantum-enhanced_2020}
L.~Banchi, Q.~Zhuang, and S.~Pirandola, Quantum-{Enhanced} {Barcode} {Decoding} and {Pattern} {Recognition}.
\newblock {\em \href{https://link.aps.org/doi/10.1103/PhysRevApplied.14.064026}{Physical Review Applied}} \href{https://link.aps.org/doi/10.1103/PhysRevApplied.14.064026}{{\bfseries 14}, 064026} (2020).

\bibitem{quantum_computational_imaging}
M.~Sarovar, {Quantum computational imaging and sensing}\href{https://doi.org/10.1117/12.2680837}{. In {\em Quantum Nanophotonic Materials, Devices, and Systems 2023}} eds.{} C.~Soci, M.~T. Sheldon, and I.~Aharonovich, \href{https://doi.org/10.1117/12.2680837}{Vol.{} 12657, 1265703} (2023).

\bibitem{debry_experimental_2023}
K.~DeBry, J.~Sinanan-Singh, C.~D. Bruzewicz, D.~Reens, M.~E. Kim, M.~P. Roychowdhury, R.~McConnell, I.~L. Chuang, and J.~Chiaverini, Experimental {Quantum} {Channel} {Discrimination} {Using} {Metastable} {States} of a {Trapped} {Ion}.
\newblock {\em \href{https://link.aps.org/doi/10.1103/PhysRevLett.131.170602}{Physical Review Letters}} \href{https://link.aps.org/doi/10.1103/PhysRevLett.131.170602}{{\bfseries 131}, 170602} (2023).

\bibitem{sinanan-singh_single-shot_2024}
J.~Sinanan-Singh, G.~L. Mintzer, I.~L. Chuang, and Y.~Liu, Single-shot {Quantum} {Signal} {Processing} {Interferometry}.
\newblock {\em \href{https://quantum-journal.org/papers/q-2024-07-30-1427/}{Quantum}} \href{https://quantum-journal.org/papers/q-2024-07-30-1427/}{{\bfseries 8}, 1427} (2024).

\bibitem{liao_quantum-enhanced_2024}
P.~Liao, B.~Zhang, and Q.~Zhuang, Quantum-enhanced learning with a controllable bosonic variational sensor network.
\newblock {\em \href{https://doi.org/10.1088/2058-9565/ad752d}{Quantum Science and Technology}} \href{https://doi.org/10.1088/2058-9565/ad752d}{{\bfseries 9}, 045040} (2024).

\bibitem{khan_quantum_2025}
S.~A. Khan, S.~Prabhu, L.~G. Wright, and P.~L. McMahon, {\em Quantum {Computational}-{Sensing} {Advantage}}.
\newblock arXiv:2507.16918 [quant-ph] (2025).

\bibitem{khan_quantum_2025-1}
S.~A. Khan, S.~Prabhu, L.~G. Wright, and P.~L. McMahon, Quantum computational sensing using quantum signal processing, quantum neural networks, and {Hamiltonian} engineering.
\newblock {\em \href{https://www.nature.com/articles/s41534-026-01235-w}{npj Quantum Information}} \href{https://www.nature.com/articles/s41534-026-01235-w}{{\bfseries 12}, 119} (2026).

\bibitem{degen2017quantum}
C.~L. Degen, F.~Reinhard, and P.~Cappellaro, Quantum sensing.
\newblock {\em \href{https://link.aps.org/doi/10.1103/RevModPhys.89.035002}{Rev. Mod. Phys.}} \href{https://link.aps.org/doi/10.1103/RevModPhys.89.035002}{{\bfseries {\bfseries 89}}3, , 035002} (2017).

\bibitem{hu_tackling_2023}
F.~Hu, G.~Angelatos, S.~A. Khan, M.~Vives, E.~Türeci, L.~Bello, G.~E. Rowlands, G.~J. Ribeill, and H.~E. Türeci, Tackling {Sampling} {Noise} in {Physical} {Systems} for {Machine} {Learning} {Applications}: {Fundamental} {Limits} and {Eigentasks}.
\newblock {\em \href{https://link.aps.org/doi/10.1103/PhysRevX.13.041020}{Physical Review X}} \href{https://link.aps.org/doi/10.1103/PhysRevX.13.041020}{{\bfseries 13}, 041020} (2023).

\bibitem{meyer_quantum_2025}
J.~J. Meyer, S.~Khatri, D.~Stilck~França, J.~Eisert, and P.~Faist, Quantum {Metrology} in the {Finite}-{Sample} {Regime}.
\newblock {\em \href{https://link.aps.org/doi/10.1103/qbn1-p6bq}{PRX Quantum}} \href{https://link.aps.org/doi/10.1103/qbn1-p6bq}{{\bfseries 6}, 030336} (2025).

\bibitem{low_optimal_hamiltonian_sim}
G.~H. Low and I.~L. Chuang, Optimal Hamiltonian Simulation by Quantum Signal Processing.
\newblock {\em \href{https://link.aps.org/doi/10.1103/PhysRevLett.118.010501}{Phys. Rev. Lett.}} \href{https://link.aps.org/doi/10.1103/PhysRevLett.118.010501}{{\bfseries {\bfseries 118}}1, , 010501} (2017).

\bibitem{low_composite_quantum_gates}
G.~H. Low, T.~J. Yoder, and I.~L. Chuang, Methodology of Resonant Equiangular Composite Quantum Gates.
\newblock {\em \href{https://link.aps.org/doi/10.1103/PhysRevX.6.041067}{Phys. Rev. X}} \href{https://link.aps.org/doi/10.1103/PhysRevX.6.041067}{{\bfseries {\bfseries 6}}4, , 041067} (2016).

\bibitem{martyn2021grand}
J.~M. Martyn, Z.~M. Rossi, A.~K. Tan, and I.~L. Chuang, Grand Unification of Quantum Algorithms.
\newblock {\em \href{https://link.aps.org/doi/10.1103/PRXQuantum.2.040203}{PRX Quantum}} \href{https://link.aps.org/doi/10.1103/PRXQuantum.2.040203}{{\bfseries {\bfseries 2}}4, , 040203} (2021).

\bibitem{liu2025toward}
Y.~Liu, J.~M. Martyn, J.~Sinanan-Singh, K.~C. Smith, S.~M. Girvin, and I.~L. Chuang, Toward Mixed Analog-Digital Quantum Signal Processing: Quantum AD/DA Conversion and the Fourier Transform.
\newblock {\em IEEE Transactions on Signal Processing} {\bfseries 73}, 3641--3655 (2025).

\bibitem{martyn2025parallel}
J.~M. Martyn, Z.~M. Rossi, K.~Z. Cheng, Y.~Liu, and I.~L. Chuang, Parallel quantum signal processing via polynomial factorization.
\newblock {\em Quantum} {\bfseries 9}, 1834 (2025).

\bibitem{allen_quantum_2025}
R.~R. {Allen}, F.~{Machado}, I.~L. {Chuang}, H.-Y. {Huang}, and S.~{Choi}, {Quantum Computing Enhanced Sensing}.
\newblock {\em arXiv e-prints} p. arXiv:2501.07625 (2025).

\bibitem{prabhu_khan_QCDS_2026}
S.~{Prabhu}, S.~A. {Khan}, X.~{Song}, M.~{Ouellet}, R.~{Yanagimoto}, S.~{Roy}, A.~{Senanian}, L.~G. {Wright}, V.~{Fatemi}, and P.~L. {McMahon}, {Quantum computational displacement sensing}.
\newblock {\em arXiv e-prints} p. arXiv:2604.13177 (2026).

\bibitem{krantz_quantum_2019}
P.~Krantz, M.~Kjaergaard, F.~Yan, T.~P. Orlando, S.~Gustavsson, and W.~D. Oliver, A {Quantum} {Engineer}'s {Guide} to {Superconducting} {Qubits}.
\newblock {\em \href{http://arxiv.org/abs/1904.06560}{Applied Physics Reviews}} \href{http://arxiv.org/abs/1904.06560}{{\bfseries 6}, 021318} (2019).

\bibitem{flux_tunable_transmon}
J.~Koch, T.~M. Yu, J.~Gambetta, A.~A. Houck, D.~I. Schuster, J.~Majer, A.~Blais, M.~H. Devoret, S.~M. Girvin, and R.~J. Schoelkopf, Charge-insensitive qubit design derived from the Cooper pair box.
\newblock {\em \href{https://link.aps.org/doi/10.1103/PhysRevA.76.042319}{Phys. Rev. A}} \href{https://link.aps.org/doi/10.1103/PhysRevA.76.042319}{{\bfseries {\bfseries 76}}4, , 042319} (2007).

\bibitem{kristen_amplitude_2020}
M.~Kristen, A.~Schneider, A.~Stehli, T.~Wolz, S.~Danilin, H.~S. Ku, J.~Long, X.~Wu, R.~Lake, D.~P. Pappas, A.~V. Ustinov, and M.~Weides, Amplitude and frequency sensing of microwave fields with a superconducting transmon qudit.
\newblock {\em \href{https://www.nature.com/articles/s41534-020-00287-w}{npj Quantum Information}} \href{https://www.nature.com/articles/s41534-020-00287-w}{{\bfseries 6}, 57} (2020).

\bibitem{taylor_high-sensitivity_2008}
J.~M. Taylor, P.~Cappellaro, L.~Childress, L.~Jiang, D.~Budker, P.~R. Hemmer, A.~Yacoby, R.~Walsworth, and M.~D. Lukin, High-sensitivity diamond magnetometer with nanoscale resolution.
\newblock {\em \href{https://www.nature.com/articles/nphys1075}{Nature Physics}} \href{https://www.nature.com/articles/nphys1075}{{\bfseries 4}, 810--816} (2008).

\bibitem{de_lange_single-spin_2011}
G.~de~Lange, D.~Ristè, V.~V. Dobrovitski, and R.~Hanson, Single-{Spin} {Magnetometry} with {Multipulse} {Sensing} {Sequences}.
\newblock {\em \href{https://link.aps.org/doi/10.1103/PhysRevLett.106.080802}{Physical Review Letters}} \href{https://link.aps.org/doi/10.1103/PhysRevLett.106.080802}{{\bfseries 106}, 080802} (2011).

\bibitem{naydenov_dynamical_2011}
B.~Naydenov, F.~Dolde, L.~T. Hall, C.~Shin, H.~Fedder, L.~C.~L. Hollenberg, F.~Jelezko, and J.~Wrachtrup, Dynamical decoupling of a single-electron spin at room temperature.
\newblock {\em \href{https://link.aps.org/doi/10.1103/PhysRevB.83.081201}{Physical Review B}} \href{https://link.aps.org/doi/10.1103/PhysRevB.83.081201}{{\bfseries 83}, 081201} (2011).

\bibitem{hall_ultrasensitive_2010}
L.~T. Hall, C.~D. Hill, J.~H. Cole, and L.~C.~L. Hollenberg, Ultrasensitive diamond magnetometry using optimal dynamic decoupling.
\newblock {\em \href{https://link.aps.org/doi/10.1103/PhysRevB.82.045208}{Physical Review B}} \href{https://link.aps.org/doi/10.1103/PhysRevB.82.045208}{{\bfseries 82}, 045208} (2010).

\bibitem{maze_nanoscale_2008}
J.~R. Maze, P.~L. Stanwix, J.~S. Hodges, S.~Hong, J.~M. Taylor, P.~Cappellaro, L.~Jiang, M.~V.~G. Dutt, E.~Togan, A.~S. Zibrov, A.~Yacoby, R.~L. Walsworth, and M.~D. Lukin, Nanoscale magnetic sensing with an individual electronic spin in diamond.
\newblock {\em \href{https://www.nature.com/articles/nature07279}{Nature}} \href{https://www.nature.com/articles/nature07279}{{\bfseries 455}, 644--647} (2008).

\bibitem{dobvsivcek2007arbitrary}
M.~Dob\ifmmode \check{s}\else \v{s}\fi{}\'{\i}\ifmmode~\check{c}\else \v{c}\fi{}ek, G.~Johansson, V.~Shumeiko, and G.~Wendin, Arbitrary accuracy iterative quantum phase estimation algorithm using a single ancillary qubit: A two-qubit benchmark.
\newblock {\em \href{https://link.aps.org/doi/10.1103/PhysRevA.76.030306}{Phys. Rev. A}} \href{https://link.aps.org/doi/10.1103/PhysRevA.76.030306}{{\bfseries {\bfseries 76}}3, , 030306(R)} (2007).

\bibitem{giovannetti_quantum-enhanced_2004}
V.~Giovannetti, S.~Lloyd, and L.~Maccone, Quantum-{Enhanced} {Measurements}: {Beating} the {Standard} {Quantum} {Limit}.
\newblock {\em \href{https://www.science.org/doi/10.1126/science.1104149}{Science}} \href{https://www.science.org/doi/10.1126/science.1104149}{{\bfseries 306}, 1330--1336} (2004).

\bibitem{giovannetti_advances_2011}
V.~Giovannetti, S.~Lloyd, and L.~Maccone, Advances in quantum metrology.
\newblock {\em \href{https://www.nature.com/articles/nphoton.2011.35}{Nature Photonics}} \href{https://www.nature.com/articles/nphoton.2011.35}{{\bfseries 5}, 222--229} (2011).

\bibitem{schuld_circuit-centric_2020}
M.~Schuld, A.~Bocharov, K.~M. Svore, and N.~Wiebe, Circuit-centric quantum classifiers.
\newblock {\em \href{https://link.aps.org/doi/10.1103/PhysRevA.101.032308}{Physical Review A}} \href{https://link.aps.org/doi/10.1103/PhysRevA.101.032308}{{\bfseries 101}, 032308} (2020).

\bibitem{grochowski_dist_phase_insens_sensing}
P.~T. {Grochowski}, M.~{Fadel}, and R.~{Filip}, {Distributed Phase-Insensitive Displacement Sensing}.
\newblock {\em arXiv e-prints} p. arXiv:2602.03727 (2026).

\bibitem{isogawa_entanglement-assisted_2026}
T.~Isogawa, G.~Wang, B.~Li, Z.~Hu, S.~Nishimura, A.~Kanamoto, H.~Yuan, and P.~Cappellaro, Entanglement-{Assisted} {Multiparameter} {Estimation} with a {Solid}-{State} {Quantum} {Sensor}.
\newblock {\em \href{https://link.aps.org/doi/10.1103/kqfr-bbfx}{PRX Quantum}} \href{https://link.aps.org/doi/10.1103/kqfr-bbfx}{{\bfseries 7}, 020307} (2026).

\bibitem{aslam_quantum_2023}
N.~Aslam, H.~Zhou, E.~K. Urbach, M.~J. Turner, R.~L. Walsworth, M.~D. Lukin, and H.~Park, Quantum sensors for biomedical applications.
\newblock {\em \href{https://www.nature.com/articles/s42254-023-00558-3}{Nature Reviews Physics}} \href{https://www.nature.com/articles/s42254-023-00558-3}{{\bfseries 5}, 157--169} (2023).

\bibitem{jin_four_2026}
X.~Jin, P.~Srivastava, R.~Wang, Y.~Li, J.~Beaumariage, T.~Purdy, M.~V.~G. Dutt, K.~Kim, K.~Seshadreesan, and J.~Liu, Four {Generations} of {Quantum} {Biomedical} {Sensors}.
\newblock {\em \href{http://arxiv.org/abs/2603.29944}{arXiv}} (2026).
\newblock arXiv:2603.29944 [quant-ph].

\bibitem{taylor_biological_2013}
M.~A. Taylor, J.~Janousek, V.~Daria, J.~Knittel, B.~Hage, H.-A. Bachor, and W.~P. Bowen, Biological measurement beyond the quantum limit.
\newblock {\em \href{https://www.nature.com/articles/nphoton.2012.346}{Nature Photonics}} \href{https://www.nature.com/articles/nphoton.2012.346}{{\bfseries 7}, 229--233} (2013).

\bibitem{fagaly_superconducting_2006}
R.~L. Fagaly, Superconducting quantum interference device instruments and applications.
\newblock {\em \href{https://doi.org/10.1063/1.2354545}{Review of Scientific Instruments}} \href{https://doi.org/10.1063/1.2354545}{{\bfseries 77}, 101101} (2006).

\bibitem{gross_magnetoencephalography_2019}
J.~Gross, Magnetoencephalography in {Cognitive} {Neuroscience}: {A} {Primer}.
\newblock {\em \href{https://www.sciencedirect.com/science/article/pii/S0896627319305999}{Neuron}} \href{https://www.sciencedirect.com/science/article/pii/S0896627319305999}{{\bfseries 104}, 189--204} (2019).

\bibitem{quantum_assured_magnav}
M.~{Muradoglu}, M.~T. {Johnsson}, N.~M. {Wilson}, Y.~{Cohen}, D.~{Shin}, T.~{Navickas}, T.~{Pyragius}, D.~{Thomas}, D.~{Thompson}, S.~I. {Moore}, M.~{Tanvir Rahman}, A.~{Walker}, I.~{Dutta}, S.~{Bijjahalli}, J.~{Berlocher}, M.~R. {Hush}, R.~P. {Anderson}, S.~S. {Szigeti}, and M.~J. {Biercuk}, {Quantum-assured magnetic navigation achieves positioning accuracy better than a strategic-grade INS in airborne and ground-based field trials}.
\newblock {\em arXiv e-prints} p. arXiv:2504.08167 (2025).

\bibitem{rossi_multivariable_2022}
Z.~M. Rossi and I.~L. Chuang, Multivariable quantum signal processing ({M}-{QSP}): prophecies of the two-headed oracle.
\newblock {\em \href{https://quantum-journal.org/papers/q-2022-09-20-811/}{Quantum}} \href{https://quantum-journal.org/papers/q-2022-09-20-811/}{{\bfseries 6}, 811} (2022).

\bibitem{QSVT}
A.~Gily{\'e}n, Y.~Su, G.~H. Low, and N.~Wiebe, Quantum singular value transformation and beyond: exponential improvements for quantum matrix arithmetics\href{https://doi.org/10.1145/3313276.3316366}{\href{https://doi.org/10.1145/3313276.3316366}{. In {\em Proceedings of the 51st Annual ACM SIGACT Symposium on Theory of Computing}, STOC 2019}, New York, NY, USA, 193–204} (2019).

\bibitem{ragone_lie_2024}
M.~Ragone, B.~N. Bakalov, F.~Sauvage, A.~F. Kemper, C.~Ortiz~Marrero, M.~Larocca, and M.~Cerezo, A {Lie} algebraic theory of barren plateaus for deep parameterized quantum circuits.
\newblock {\em \href{https://www.nature.com/articles/s41467-024-49909-3}{Nature Communications}} \href{https://www.nature.com/articles/s41467-024-49909-3}{{\bfseries 15}, 7172} (2024).

\bibitem{QSP_phase_factor_eval}
Y.~Dong, X.~Meng, K.~B. Whaley, and L.~Lin, Efficient phase-factor evaluation in quantum signal processing.
\newblock {\em \href{https://link.aps.org/doi/10.1103/PhysRevA.103.042419}{Phys. Rev. A}} \href{https://link.aps.org/doi/10.1103/PhysRevA.103.042419}{{\bfseries {\bfseries 103}}4, , 042419} (2021).

\end{thebibliography}

\appendix
\setcounter{figure}{0}
\renewcommand{\thefigure}{A\arabic{figure}}

\section{Experimental Setup}
\label{app_sec:experimental_setup}

\subsection{Setup}

The device consists of a double-junction flux-tunable transmon fabricated by depositing aluminum on a silicon substrate. The qubit is capacitively coupled to a readout resonator, which is inductively coupled to a Purcell filter. The Purcell filter is capacitively coupled to the readout drive line used for reflection measurement. The qubit is capacitively coupled to the drive line and the SQUID loop is inductively coupled to an on-chip flux-bias line. The device is thermally anchored to the mixing chamber and shielded by a copper enclosure coated with Berkeley Black, with an additional outer aluminum shield. The device is also shielded from external magnetic fields by a mu-metal shield around the fridge.
\begin{figure}[htb]
    \centering
    \includegraphics[width=\linewidth]{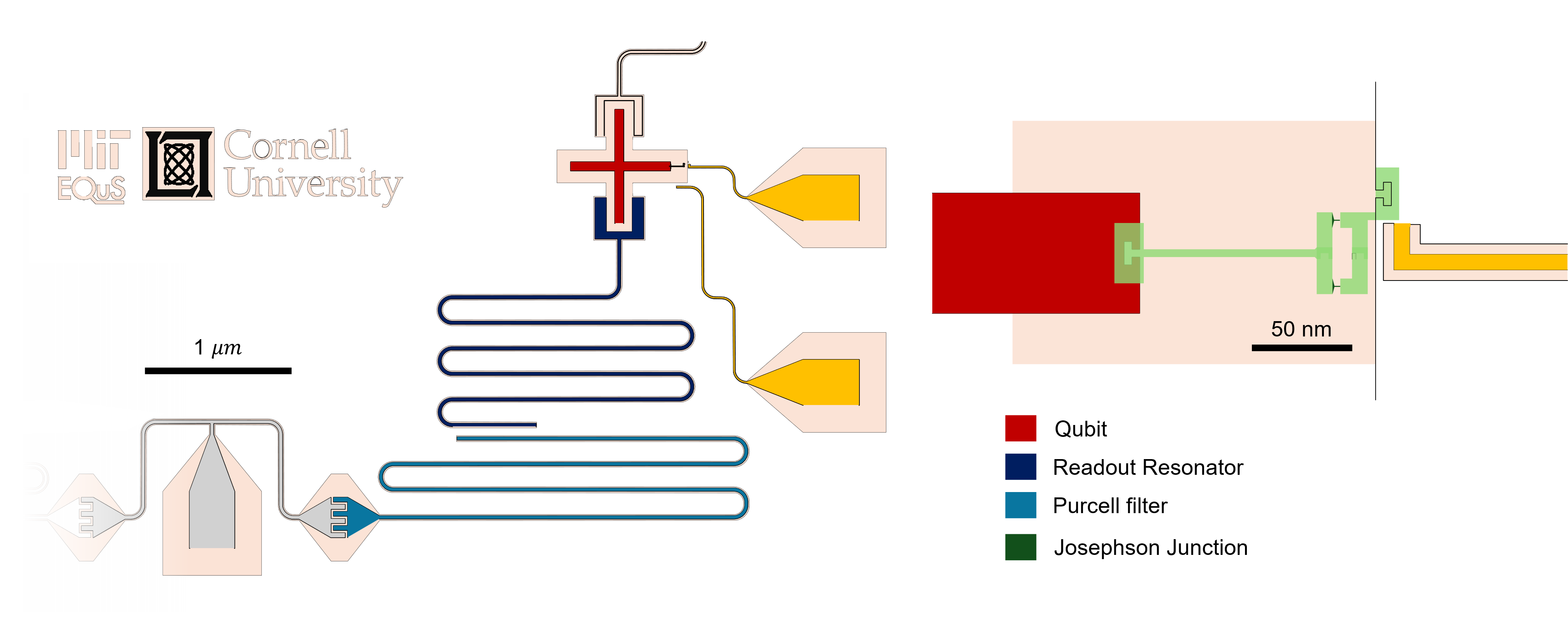}
    \caption{\textbf{False-colored optical micrograph of the superconducting quantum device used for quantum computational sensing experiments.} The device consists of a flux-tunable transmon qubit (red) capacitively coupled to a readout resonator (dark blue), which is in turn coupled to a Purcell filter (light blue) for enhanced readout and suppression of radiative decay. The qubit frequency is controlled through an on-chip flux-bias line, enabling tunable sensitivity to external magnetic flux signals. Right: false-colored Micrograph of the Josephson junction forming the nonlinear element of the transmon qubit.}
    \label{app_fig:wiring diagram}
\end{figure}
The device was cooled and measured in a \textit{Bluefors LD250} dilution refrigerator with a base temperature of approximately 9 mK. A schematic of the cryogenic setup and associated wiring is shown in Fig.~\ref{app_fig:wiring diagram}. Three input lines were used: a readout drive line, a qubit drive line, and a fast flux control line capable of applying both static and oscillating magnetic flux to the SQUID loop. The readout and qubit drive signals were generated using a \textit{Xilinx RFSoC 4X2} board with a sampling frequency of 9530.4 MHz. The qubit drive line was amplified using a \textit{Mini-Circuits ZX60-123LN+} amplifier. For the flux line, we used a bias tee outside the fridge to combine a static and oscillating signal generated by a \textit{Yokogawa GS200} and an \textit{SDG2042X} respectively. 

\begin{figure}[htb]
    \centering
    \includegraphics[width=0.7\linewidth]{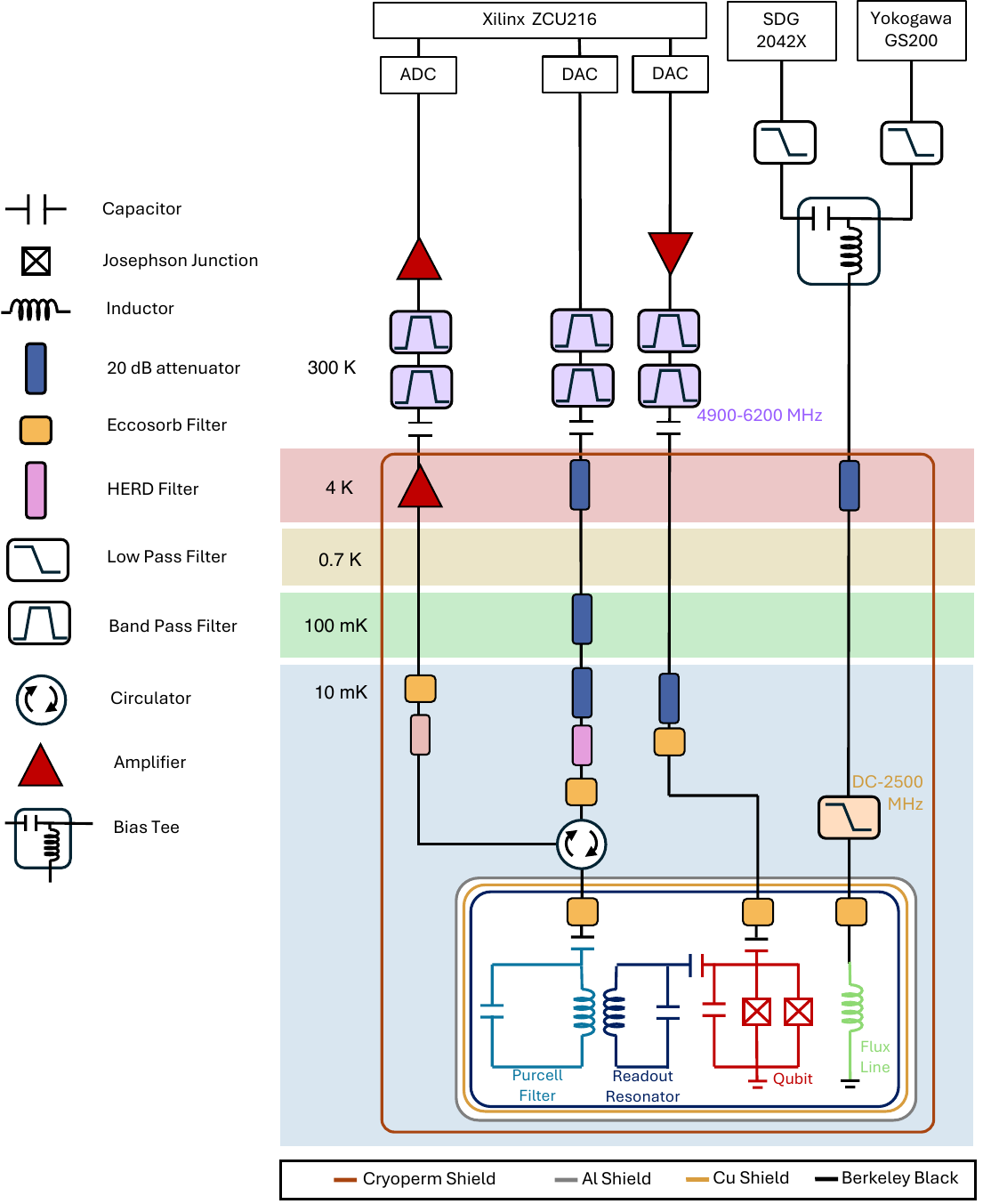}
    \caption{\textbf{Wiring diagram.}
    Schematic of the experimental setup, showing the room-temperature electronics, cryogenic input and output lines, filtering and attenuation stages, and shielding around the device inside the dilution refrigerator.}
    \label{app_fig:wiring diagram}
\end{figure}

\subsection{System Hamiltonian}
\label{app_ssec:System Hamiltonian}
For our sensing experiments, we employ an asymmetric double-junction flux-tunable transmon, in which a single Josephson junction is replaced by a DC SQUID to make the qubit frequency tunable. The system can be described by the Hamiltonian \cite{krantz_quantum_2019}.
\begin{equation}
\label{app_eq: DJT_1}
H = 4E_C n^2 -
\underbrace{E_{J\Sigma}\sqrt{\cos^2(\varphi_e) + d^2 \sin^2(\varphi_e)}}_{E'_J(\varphi_e)}
\cos(\phi),
\end{equation}
where $E_C$ is the charging energy, $E_{J\Sigma}$ is the sum of the Josephson energies of the two junctions, and $d=(\gamma-1)/(\gamma+1)$ parameterizes the junction asymmetry with $\gamma = E_{J2}/E_{J1}$. $\varphi_e=\pi \tfrac{\Phi_{ext}}{\Phi_0}$, where $\Phi_{ext}$ denotes the applied external flux with $\Phi_0$ being the superconducting magnetic flux quantum. $n$ and $\phi$ are the conjugate charge and phase operators. The flux dependence of the Hamiltonian is captured through the effective Josephson energy $E'_J(\varphi_e)$.
In the transmon regime, expanding the cosine:
\begin{equation}
\label{app_eq: DJT_2}
H \approx 4E_C n^2 - E'_J(\varphi_e)\left(1 - \frac{\phi^2}{2} + \frac{\phi^4}{24} + \mathcal{O}(\phi^6)\right),
\end{equation}
This corresponds to a weakly anharmonic oscillator with a flux-dependent transition frequency. The qubit is biased away from a flux sweet spot to enhance its sensitivity to small perturbations in the external flux. It is driven on resonance at this bias point as shown in Fig.~\ref{app_fig:calibration}, such that any flux-induced frequency shift results in a detuning of the drive. This detuning causes the qubit to accumulate phase, corresponding to a rotation about the Z-axis, resulting in an effective Z gate.

\subsection{Calibration}

\label{app_ssec:Calibration}
Here we describe the calibration procedure used to obtain the parameters listed in Table~\ref{app_tab:calibration}.

\begin{figure}[htb]
    \centering
    \includegraphics[width=\linewidth]{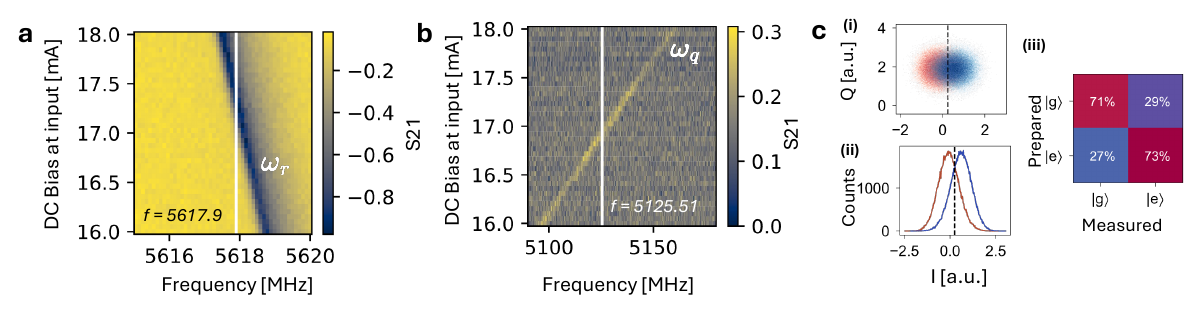}
    \caption{\textbf{ Flux-dependent qubit and readout calibration.} \textbf{a,} Readout resonance frequency as a function of applied flux. The readout frequency varies only weakly over the experimentally relevant range. \textbf{b,} Qubit transition frequency as a function of applied flux, showing a much stronger dependence than the readout mode. This difference allows a single readout frequency to be used throughout the experiment, even though the instantaneous signal, and therefore the qubit frequency, is not known at the time of measurement. \textbf{c,} Readout calibration from the rotated IQ distributions, including the assignment histogram with threshold set at 0 and the corresponding readout fidelity matrix.}
    \label{app_fig:calibration}
\end{figure}

The readout resonator frequency was determined by sweeping a weak readout pulse in frequency, recording the averaged IQ response at each point, and fitting the minimum of the measured response to extract the resonance frequency. The readout resonator frequency was then measured as a function of bias current as shown in Fig.~\ref{app_fig:calibration}a by repeating this spectroscopy procedure over a range of applied currents, extracting the resonance frequency at each point from the minimum of the measured response. The optimal readout resonator frequency is further fine-tuned to get optimal single-shot readout fidelity as shown in Fig.~\ref{app_fig:calibration}c.  

\begin{table}[t]

\centering
\renewcommand{\arraystretch}{1.15}
\setlength{\tabcolsep}{10pt}

\begin{tabular}{c|c|c!{\hspace{2pt}\vrule\hspace{2pt}}c}
\multicolumn{1}{c}{Parameter} & \multicolumn{1}{c}{Mode(s)} & \multicolumn{1}{c}{Symbol} & \multicolumn{1}{c}{Value} \\
\hline
\hline
Frequency & Transmon g-e & $\omega_q$ & $2\pi\times 5.125~\mathrm{GHz}$ \\
          & Readout      & $\omega_r$ & $2\pi\times 5.618~\mathrm{GHz}$ \\

\hline
Relaxation time & Transmon g-e & $T_1$   & 30~\si{\micro\second} \\

Dephasing time & Transmon g-e & $T_2^*$ & 8~\si{\micro\second} \\

Readout Fidelity & Ground state readout error & $\mathrm{P_{eg}}$ & $29\%$ \\
                   & Excited state readout error   & $\mathrm{P_{ge}}$ & $27\%$ \\
                   \hline
Flux sensitivity near bias point & Transmon g-e & $\partial f_{ge}/\partial I_{\mathrm{flux}}$ & $\approx 29$~\si{MHz/mA}\\
\end{tabular}
\caption{\textbf{Appendix Table 1 | System parameters and dissipation rates.}
}
\label{app_tab:calibration}
\end{table}

For qubit spectroscopy, a Gaussian-shaped amplitude modulated pulse was applied to drive the g-e transition, sweeping the carrier frequency, and the state was dispersively measured using a fixed flat-top readout pulse on the readout resonator.
The qubit spectroscopy was then repeated over a range of bias currents, updating the readout frequency at each point to match the calibrated resonator frequency (see Fig.~\ref{app_fig:calibration}b) to extract the flux dependence of the g-e transition frequency as discussed in section~\ref{app_ssec:System Hamiltonian}.

For our classification experiments, the qubit was biased away from the sweet spot to make the frequency sensitive to the applied flux and the qubit and readout frequencies were calibrated at this point to get optimal readout fidelity. 
The $\pi$-pulse was calibrated at this bias point using a Rabi experiment by sweeping the pulse amplitude. Pulses for other rotation angles were obtained by linear interpolation from this calibration. Readout assignment is performed by first rotating the measured IQ data so that the separation between the $|g\rangle$ and $|e\rangle$ responses lies along a single quadrature, and then applying a threshold on this rotated axis to classify each single-shot measurement. The same thresholding procedure also defines the readout fidelity matrix, whose entries are given by the fractions of the prepared $|g\rangle$ and $|e\rangle$ populations assigned to each outcome reported in Fig.~\ref{app_fig:calibration}c.
The $T_1$ and $T_2^*$ measurements were performed at the bias point, and the extracted values are reported in Table \ref{app_tab:calibration}. 

We perform the classification experiments over a qubit detuning range of 5 MHz around this operating point. Over this range, the readout resonator frequency shifts by only $\sim 10^{-1}$ MHz, enabling stable dispersive readout without retuning the resonator, which remains fixed while the input signal to be classified varies.

\section{Time-Independent QCS Protocol}
\label{app_sec:DC_QCS}

\subsection{Experimental Details}

A trained QSP protocol was first loaded from the saved simulation file, which contained the optimized pulse parameters and the simulated expectation-value response as a function of detuning. 
To implement the normalized classification tasks experimentally, the normalized variable $x\in[0,1]$ was mapped onto the physical signal range through the affine transformation
\begin{equation}
    u=u_{\min}+(u_{\max}-u_{\min})x.
\end{equation}
For static-field classification, this transformation mapped the normalized task boundaries onto the qubit-detuning range from $0$ to $5~\mathrm{MHz}$.

Before classification, a separate calibration run of 1000 shots was performed using the loaded protocol at 100 points spanning a broader signal range. 
This calibration was used to refine the affine mapping between the experimentally applied flux signal and the effective detuning used in the simulation. 
The resulting calibration parameters were fixed for all subsequent measurements.

For each shot, the IQ readout was rotated in the IQ plane and thresholded to obtain a binary measurement outcome. 
These outcomes were averaged over repeated shots to obtain an empirical excitation probability.
When required for calibration or comparison with simulation, the averaged response was further corrected using the readout-fidelity matrix.

For all classification datasets, candidate signal values were sampled uniformly over the full operating range and assigned labels according to the task boundaries. 
Rejection sampling was then used to retain equal numbers of values from the two classes. 
To evaluate performance for different shot budgets, a training dataset was acquired at 100 signal values sampled using this procedure, with 1000 shots collected at each value. Subsets of these shots were used to emulate the desired shot budget. 
A separate random-forest classifier was trained for each shot number to map the averaged experimental response onto the corresponding class label.
This classifier was chosen for simplicity and to provide a uniform analysis across all tasks and protocols. 
In nearly all cases, the resulting decision rule reduced effectively to a single threshold on the averaged response, as expected from the responses shown in the main text.

Finally, a third independent experimental run was performed using the fixed calibration parameters. This run produced independent balanced validation and test sets, each containing 50 signal values with 1000 shots per value.

\subsection{Training Details}

\subsubsection{Numerical Model}

We implement the sensor as a custom fully analog PyTorch model describing a single qubit initialized in $|0\rangle$ and subjected to a sequence of trainable $X$ rotations interleaved with sensing intervals~(with trainable rotations not commuting with the sensing unitaries). 
Each control pulse is discretized in time and applied with a Gaussian envelope whose integral is normalized to match the target rotation angle, while the sensing periods correspond to free evolution with zero drive. 
The input signal enters as a time-dependent detuning and may vary from shot to shot in the stochastic setting.

Each trainable $X$ rotation is applied using a Gaussian pulse envelope, normalized so that its time integral matches the target rotation amplitude.
The sensing intervals correspond to free evolution, implemented with a zero drive amplitude over the sensing window. 

The final state is measured in the computational basis. Depending on the setting, the model can either return the full measurement probabilities directly or emulate finite-shot readout through sampling. 
In the multi-shot setting, different shots may be propagated under different signal realizations and then either kept as sampled outcomes or averaged to return the mean expectation value.

\begin{figure}[htb]
    \centering
    \includegraphics[width=0.8\linewidth]{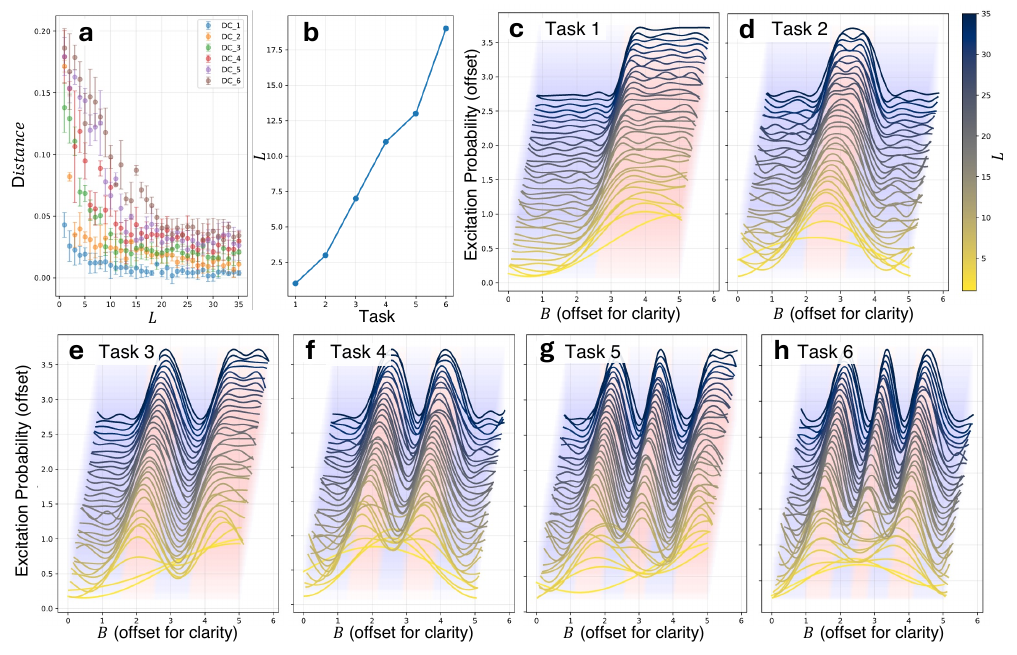}
    \caption{\textbf{Simulated performance and complexity of time-independent QCS protocols across all tasks.} \textbf{a,} Mean squared distance between the target function and the simulated learned QSP response as a function of the number of layers $L$ for Tasks 1 to 6. In all cases, the distance decreases with increasing protocol depth before approaching a plateau. \textbf{b,} Minimum number of gates required for each task to reach a distance below 0.05, showing an approximately linear increase in protocol complexity with task difficulty. \textbf{Task 1 to Task 6,} Simulated learned QSP response curves over the full range of time-independent signal amplitudes for increasing $L$. As the protocol depth increases, the response becomes more structured and progressively better matches the target function, consistent with the convergence behavior shown in panel \textbf{a}.}
    \label{app_fig:DC_training}
\end{figure}

\subsubsection{Gates Training}

For circuit depths $L\in[1,35]$, the protocol angles were initialized with a $\pi/2$ rotation at the beginning and end of the sequence, while the $L-1$ intermediate angles were initialized randomly in $[0,\pi]$. 
The phase was sampled uniformly over $[0,2\pi]$. 
Therefore, each protocol contains $L+1$ gates, where the final gate defines the measurement axis and is not counted in $L$.
For each pair $(L,S)$, we trained a QSP model using Adam with a learning rate linearly decaying from $10^{-3}$ to $10^{-5}$. 

Training data were generated on the fly by first sampling candidate signal values $u$ uniformly over the full signal range. 
Each value was assigned a binary label according to the parity of the interval defined by the task cutoffs. 
Rejection sampling was then used to retain equal numbers of samples from the two classes. 
Validation data were generated independently using the same procedure.

Two training modes were used. In sampled mode, the model output included finite-shot measurement noise through an explicit sampler. In exact mode, the sampler was disabled and the model returned the exact expectation value of the measurement outcome. Exact mode therefore corresponds to noiseless optimization of the underlying quantum response rather than of sampled projective outcomes. Empirically, this mode yielded more reliable training at low shot number, likely because it avoided the straight-through estimator and improved the gradient signal.

Validation was performed at the end of each epoch on an independently generated dataset drawn from the same distribution. Each training batch contained 64 sampled amplitude values, with 32 batches per epoch, and all models were trained for 50 epochs. To simulate the continuous dynamics of the protocol, both gate and sensing intervals were discretized into 1 \si{n s} segments.

Figure~\ref{app_fig:DC_training}a shows the mean squared distance between the target function and the trained protocol as a function of the number of layers $L$ for all six tasks. The overall decrease in distance with increasing $L$ indicates that the training successfully improves the learned response as the protocol becomes deeper. For the more difficult tasks, the improvement eventually saturates. We hypothesize that this plateau arises from the increasingly complex optimization landscape, where a larger number of local minima makes the final performance more sensitive to the initialization.

We further quantify the increase in task complexity by plotting the minimum number of layers required to achieve a loss below 0.05, as shown in Fig.~\ref{app_fig:DC_training}b. 
This threshold depth increases approximately linearly with task number, consistent with the expectation that more complex target functions require a larger number of effective Fourier components to be reproduced accurately. 
The simplest tasks can be implemented with very shallow protocols. 
In particular, Task 1 can already be approximated with a single gate (plus the final readout gate), as expected from its close similarity to a Ramsey-like response. 
Task 2 remains relatively simple, but this low-depth approximation rapidly breaks down for the more complex tasks, as shown in Fig.~\ref{app_fig:DC_training}: task 3-6.

\begin{figure}[htb]
    \centering
    \includegraphics[width=\linewidth]{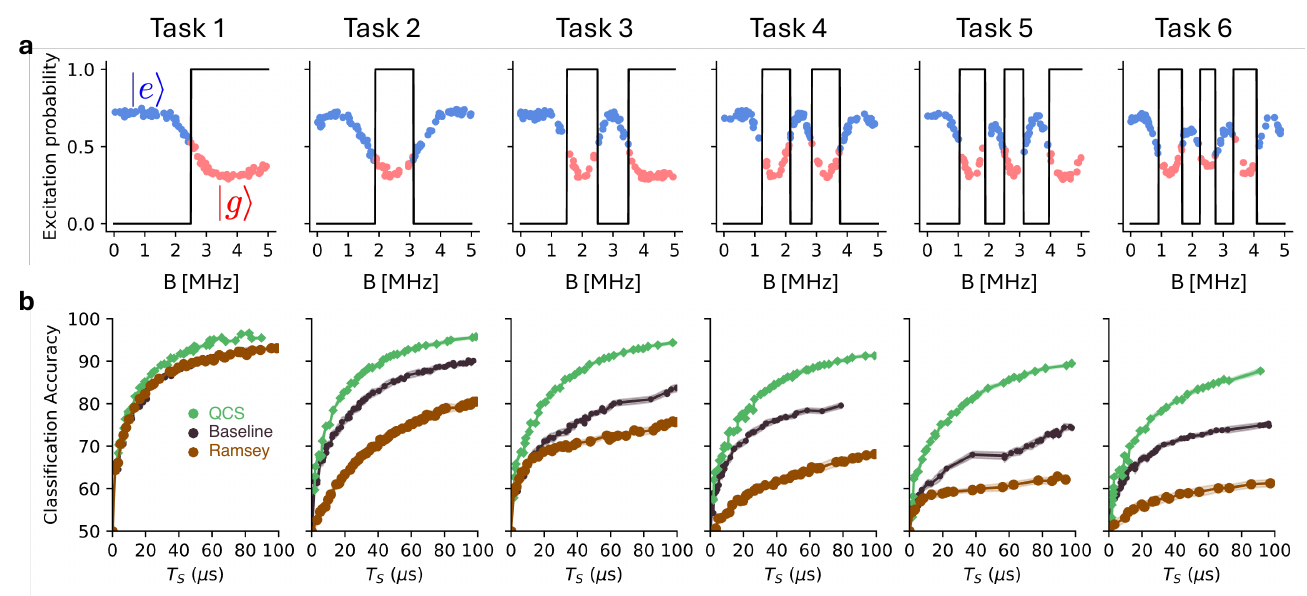}
    \caption{\textbf{  Experimental time-independent QCS performance across classification tasks.} Each column corresponds to one of the six classification tasks. \textbf{a,} learned excitation probabilities $(P_e)$ as a function of the signal, shown together with the target classification function. \textbf{b,} classification accuracy as a function of the total sensing time $(T_s)$ for the learned QSP protocol and for the conventional sensing baseline (Refer to Appendix~\ref{app_sec:DC_baseline}), along with a Ramsey protocol. In all tasks, increasing protocol duration improves the classification accuracy, while the learned QSP protocol consistently outperforms the conventional baseline over the same sensing time range.}
    \label{app_fig:dc_all_task}
\end{figure}

\subsection{Time-Independent Results Across All Tasks}

Figure~\ref{app_fig:dc_all_task} summarizes the experimental performance of the time-independent QCS protocol across all six classification tasks. For DC classification tasks, 14 QSP protocols for task 1 and 15 protocols for tasks 2 to 6 were tested, in order to probe a wider range of circuit depths.
The top row shows the learned response as a function of the signal, together with the corresponding target function for each task. For plotting, we show the response corresponding to the QSP protocol that has the maximum number of points on the Pareto front of classification accuracies as shown in the bottom row.
As the task complexity increases, the target function requires progressively sharper and more complex features in the measured response. 

The bottom row compares the classification accuracy of the learned Quantum Signal Processing (QSP) protocol with that of our conventional sensing baseline (Refer to Appendix~\ref{app_sec:DC_baseline}) as a function of the total sensing time $T_s$.
Longer sensing times consistently yield higher classification accuracy, reflecting increased information about the static field. The plotted points represent the Pareto front, capturing the optimal trade-off between accuracy and sensing time. Across all six tasks, the QSP-based protocol outperforms our baseline over the same sensing-time range, with the advantage becoming particularly clear for the more complex classification functions.

\section{Time-Independent Conventional Baseline}
\label{app_sec:DC_baseline}

\begin{figure}[htb]
    \centering
    \includegraphics[width=0.8\linewidth]{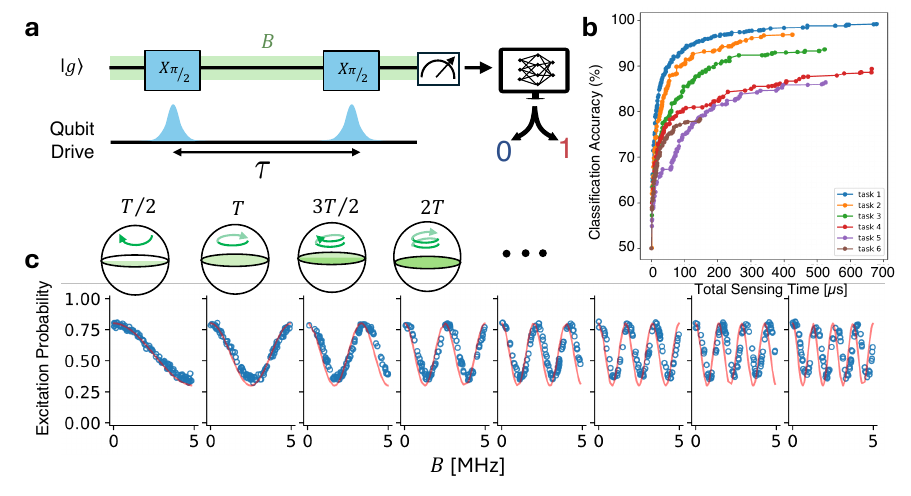}
    \caption{\textbf{Experimental realization of our conventional sensing baseline} \textbf{a,} Schematic of a single sensing step with interrogation time $\tau$, consisting of two $X_{\pi/2}$ pulses separated by free evolution under the static signal, followed by measurement and classical postprocessing for binary classification. \textbf{b,} Post-optimization test-set accuracy in the experiment for the different classification tasks, shown as a function of the total sensing time. For each task, the maximum protocol duration was chosen from the Pareto front and is limited either by the point at which further gains become negligible or by the experimental constraint of at most 96 shots per sub-protocol. \textbf{c,} Experimentally measured averaged responses for a set of interrogation times spanning eight half-period increments, from $T/2$ to $4T$. Blue markers show the measured averages, and red curves show the ideal Ramsey responses.}
    \label{app_fig:dc_baseline_exp}
\end{figure}

\subsection{Experimental Details}

For our conventional baseline, the protocol is implemented in a form closely related to the time-independent QCS protocol. 
Each sub-protocol consists of two $\pi/2$ pulses about the $X$ axis separated by a free-evolution interval, as shown in Fig.~\ref{app_fig:dc_baseline_exp}a. 
The waiting time is chosen as an integer multiple of half the time period (here, defined as $T$) of the Ramsey response in detuning space. 
We consider the first eight such waiting times, corresponding to evolutions ranging from half of a cosine cycle to four full cycles, as shown in Fig.~\ref{app_fig:dc_baseline_exp}c. 
At longer waiting times, small discrepancies begin to appear.

The eight sub-protocols can then be combined with different shot allocations, depending on which interrogation times are most informative for the classification task. Since our goal is classification rather than parameter estimation, we optimize the shot allocation across the eight Ramsey sub-protocols for each task, yielding improved performance as shown in the next section.

Three independent datasets were acquired for the DC-field classification task. 
The training set contained 100 randomly sampled signals with 1000 shots per amplitude, whereas the validation and test sets each contained 50 randomly sampled amplitudes with 1000 shots per amplitude.

Figure \ref{app_fig:dc_baseline_exp}b shows the full Pareto front obtained for time allocations up to a total sensing time of 700~\si{\micro\second}. 
Not all optimized protocols extend to this maximum duration. This behavior arises from the combination of two constraints: the cap of 96 shots per interrogation time and the diminishing accuracy gains obtained by adding longer wait times to the protocol. 
Once the improvement becomes too small, extending the protocol is no longer favorable within the allowed shot budget. 
This effect is especially pronounced for Task 6, whose optimized protocol saturates unusually early. 
The origin of this behavior is discussed in more detail in the following training section.

\begin{figure}[htb]
    \centering
    \includegraphics[width=\linewidth]{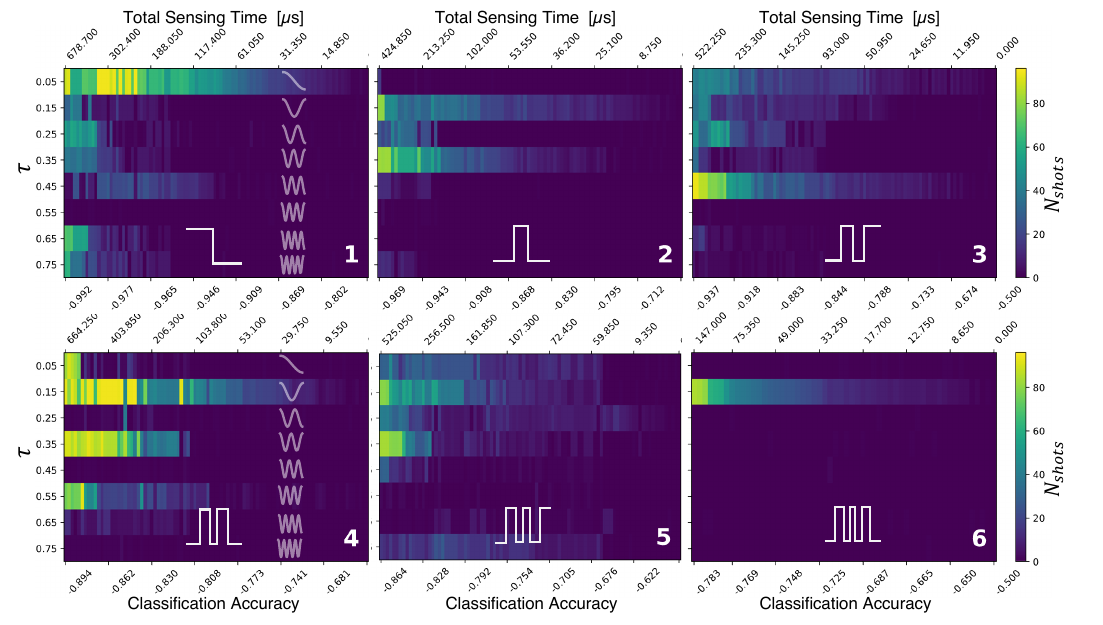}
    \caption{\textbf{Optimized Ramsey shot allocation across all static-field classification tasks.} Each panel shows the optimized distribution of shots across the eight interrogation times used for one of the six tasks, indicated by the inset target function and panel label. The color scale gives the number of shots assigned to each interrogation time, from 0 to 96. The horizontal axis shows the achieved classification accuracy, and the upper axis shows the corresponding total sensing time. These optimized shot distributions illustrate how the experimental resources are redistributed across interrogation times depending on the structure and difficulty of the target classification task.}
    \label{app_fig:dc_baseline_training}
\end{figure}

\subsection{Training Details}

To identify efficient protocols, we used NSGA-II to approximate the Pareto front defined by two competing objectives: maximizing the classification accuracy under a balanced 50:50 class prior and minimizing the total sensing time. Each candidate protocol was encoded as an 8-dimensional integer vector specifying the number of shots assigned to each of eight sensing times between 0.05 and 0.75 \si{\micro\second}, with each entry bounded between 0 and 96.

The evolutionary optimization used a population of 200 individuals over 100 generations, which we found sufficient for reliable convergence. Simulated binary crossover (SBX) and polynomial mutation (PM) were used as variation operators.
For each shot-allocation vector, a random forest classifier was trained on the training set and its performance was evaluated on the validation set. This validation accuracy served as the optimization metric used by NSGA-II to construct the Pareto front.

To estimate final performance, selected protocols were evaluated on the independent test set over 20 repeated runs, from which we report the mean accuracy and standard deviation. In each run, shots were drawn without replacement from the 1000 available shots for each amplitude.
Many solutions on the raw Pareto front exceeded the protocol-duration range of interest and were therefore removed from the displayed front. For simplicity, no explicit global constraint was imposed on the total duration of the full shot-allocation vector during optimization.

Figure \ref{app_fig:dc_baseline_training} shows the optimized shot distributions over the eight interrogation times for the full range of allowed total sensing times. 
The Pareto front is formed by a continuous evolution of these shot allocations. 
For each task, once the most informative interrogation times reach the maximum of 96 shots, additional wait times are introduced to obtain further gains in accuracy.

For the first four tasks, the optimized protocols show a clear alternation between odd and even half-period interrogation times, consistent with the structure of the target functions. 
The last two tasks mark the onset of significantly greater complexity. 
Task 5 displays a mixed use of both odd and even interrogation times, which likely underlies the change in performance behavior discussed in Fig.~\ref{app_fig:dc_baseline_exp}b. 
Task 6 instead appears to bypass the full complexity of the target by approximating the central three-feature structure as a single broader feature, leading to rapid initial improvement but only limited gains at longer sensing times, as seen in Fig.~\ref{app_fig:dc_baseline_exp}b.

\section{Time-Dependent QCS Protocol}
\label{app_sec:AC_QCS}

\begin{figure}[p]
    \centering
    \includegraphics[width=0.8\linewidth]{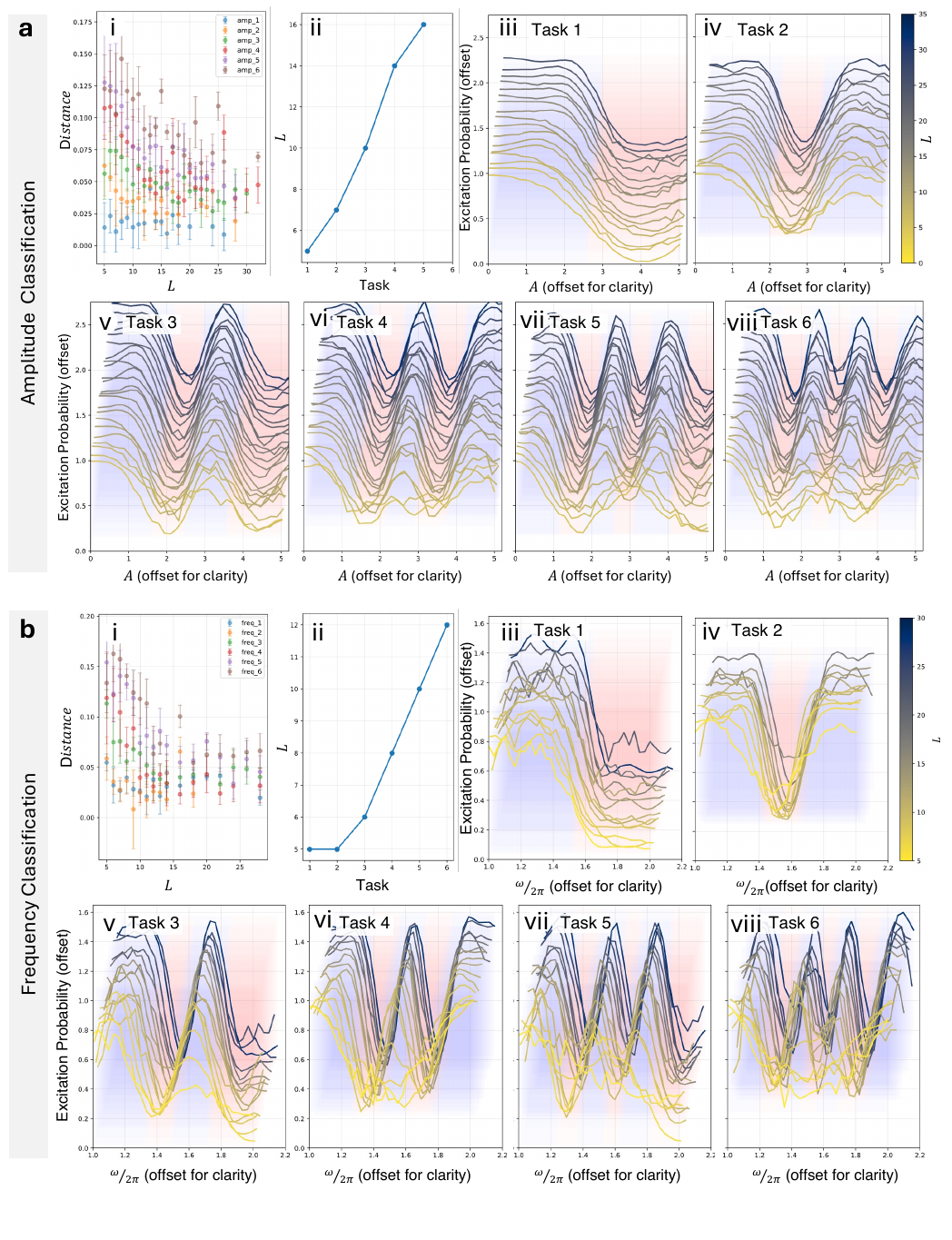}
    \caption{\textbf{Simulated performance and complexity of time-dependent QCS protocols across all tasks.}
    \textbf{a}, amplitude-classification tasks. \textbf{i}, Mean squared distance between the target function and the learned QSP response versus the number of layers $L$ for Tasks 1 to 6. \textbf{ii}, Minimum number of gates required for each task to reach a distance below 0.05. \textbf{iii-viii}, Simulated learned QSP response over the full range of time-dependent signal amplitudes for increasing $L$ for Tasks 1 to 6.
    \textbf{b}, frequency-classification tasks. \textbf{i}, Mean squared distance between the target function and the learned QSP response versus the number of layers $L$ for Tasks 1 to 6. \textbf{ii}, Minimum number of gates required for each task to reach a distance below 0.05. \textbf{iii-viii}, Simulated learned QSP response over the full range of time-dependent signal frequencies for increasing $L$ for Tasks 1 to 6.}
    \label{app_fig:supp_ac_all}
\end{figure}

\subsection{Experimental Details}

A trained QSP protocol was first loaded from the saved simulation file, which contains the optimized pulse parameters together with the simulated response as a function of AC signal amplitude/frequency. 
The experimental procedure followed the same general structure as for the DC protocol described in Sec.
\ref{app_sec:DC_QCS}, with the main difference being that the signal is now time-dependent and carries a randomized phase relative to our protocol. This stochasticity in phase was enforced by implementing a randomized wait-time between shots. As in the DC case, the protocol was first calibrated experimentally before classification. 
A separate calibration run of 1000 shots was performed for the amplitude-classification over 100 values of the applied AC drive amplitude, spanning a range broader than that of the target task. No calibration was required for the frequency-classification, as the flux frequency does not directly map onto the qubit detuning. During this experiment, the qubit was biased to its operating point using a fixed DC current, while the AC signal was generated as a sinusoidal current drive at fixed frequency with variable amplitude or frequency which were combined with a room-temperature bias tee and then sent into the fridge through an RF input line. 
Because the drive was not phase locked to the pulse sequence, its initial phase was effectively random from shot to shot. 
The calibration step was used to determine the mapping between the experimentally applied AC amplitude/frequency and the effective detuning amplitude used in simulation. 
As in Sec.~\ref{app_sec:DC_QCS}, this was done by fitting a scale factor that minimizes the mean squared difference between the corrected excitation probability and the simulated expectation curve. 
This procedure compensates for uncertainty in the amplitude-to-detuning conversion. 
Once determined, this scale factor was fixed and reused for all subsequent measurements of that protocol.
The readout procedure was identical to that used for the DC protocol in Sec.~\ref{app_sec:DC_QCS}. 

\begin{figure}
    \centering
    \includegraphics[width=\linewidth]{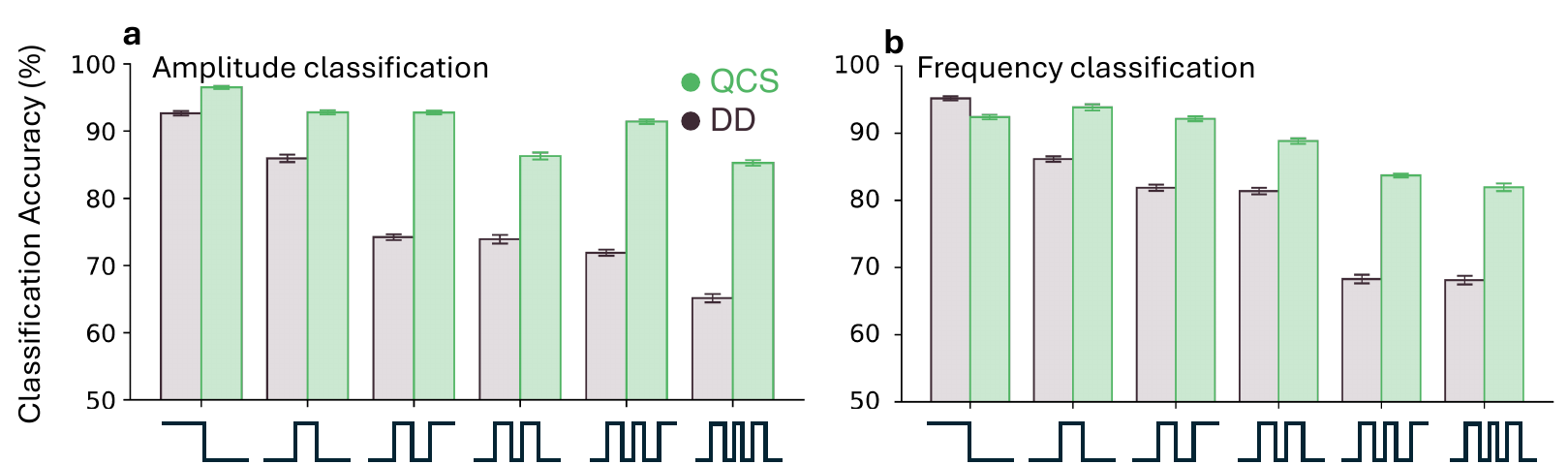}

    \vspace{1em}

    \caption{\textbf{QCS advantage for time-dependent signals}. \textbf{a}, Classification accuracies at a total sensing time of 200~\si{\micro\second} for the six time-dependent signal amplitude-classification tasks, illustrating increasing task complexity and the advantage of QCS over dynamical decoupling. \textbf{b}, Corresponding classification accuracies for frequency-classification tasks.}
    \label{app_fig:ac_am_hist}
\end{figure}
To quantify the performance of a protocol for a given number of shots, the same approach as in Sec. \ref{app_sec:DC_QCS} was used. 
In contrast to the DC protocol, the random initial phase of the AC signal introduces an additional source of stochasticity. The measured response at fixed amplitude therefore reflects not only projection noise and readout noise, but also an average over the phase realizations sampled during the experiment.

\begin{figure}
    \centering
    \includegraphics[width=\linewidth]{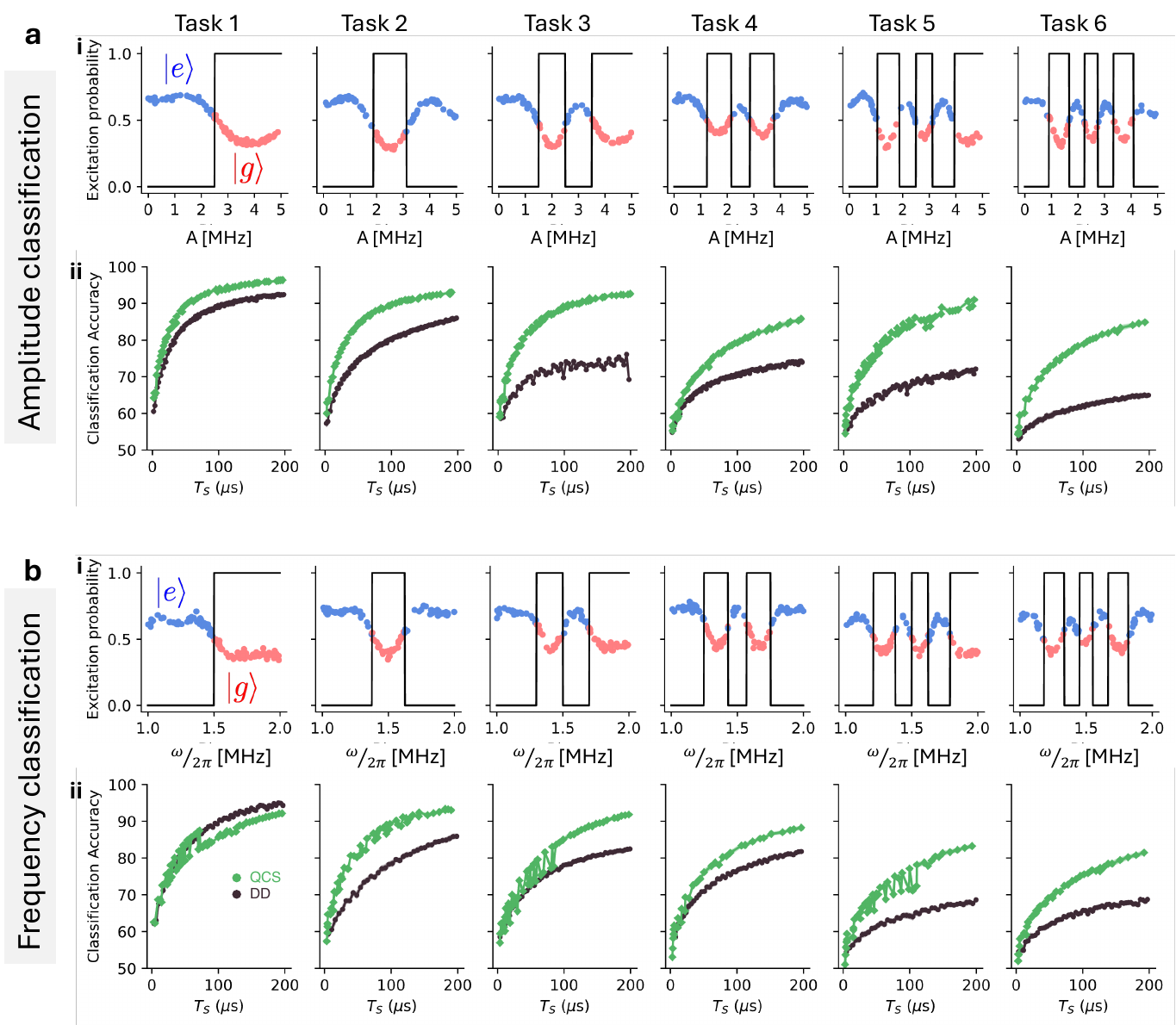}
    \caption{\textbf{Experimental time-dependent QCS performance across classification tasks.} Each column corresponds to one of the six classification tasks. \textbf{a,} \textbf{i,} learned excitation probability $P_e$ as a function of the time-dependent signal amplitude, shown together with the target classification function. \textbf{ii,} classification accuracy as a function of the total sensing time $T_s$ for the learned QSP protocol and for the optimized dynamical-decoupling-based protocol.\\ \textbf{b,} frequency-classification tasks. \textbf{i,} learned excitation probability $P_e$ as a function of the time-dependent signal frequency, shown together with the target classification function. \textbf{ii,} classification accuracy as a function of the total sensing time $T_s$ for the learned QSP protocol and for the optimized dynamical-decoupling-based protocol. In all tasks, longer protocol durations lead to higher classification accuracy. Across the same sensing-time range, the learned QSP protocol consistently outperforms the optimized DD protocol except on the simplest task, with the advantage growing as the task complexity increases.}
    \label{app_fig:ac_all_task}
\end{figure}

\subsection{Training Details}

For time-dependent signals, the training procedure is similar to that used for time independent signals, but differs in the sampling of the initial protocol and in the treatment of the phase stochasticity.
In AC sensing, the signal phase is not fixed. As a result, for each simulated amplitude or frequency, the protocol must be evaluated repeatedly over different phase values.
To account for this, we simulated each shot individually. 
To account for phase averaging of the excitation probability across shots, we selected a fixed number of shots (32), each with a randomly sampled phase, and averaged the corresponding expectation values to obtain the effective measured response.
This procedure increases the computational cost and introduces stochasticity into the gradients, even when the exact expectation value is used rather than sampled projective outcomes.
Because of the variability in the initial phase of the incoming magnetic field, we expected the optimization landscape to be highly non-smooth, making optimization difficult. 
To mitigate this, we first constructed a database of simulated protocols to seed the optimization containing $199186$ candidates for amplitude-classification and $150283$ candidates for frequency-classification.
Initial guesses were generated using two complementary random initialization schemes. In the first scheme, the protocol was sampled fully at random, as in the time independent case, except that in half of the initializations, all rotation angles were set to zero.
For the second scheme, the initial guess was chosen to resemble a dynamical-decoupling-like sequence. A random walk over rational multiples of $\pi$ was used to generate a more structured set of gate angles. As in the fully random case, the additional rotation angles were either set to zero or sampled randomly.

For a given task, we then identified 100 candidate protocols from the Pareto front defined by the distance between the averaged expectation value and the target function. 
These candidates were subsequently filtered according to the maximum allowed sequence length.
Each retained Pareto-optimal point was then used as an initialization for a further optimization run, following the same procedure used for the time independent task. 
Finally, the best-performing protocols were selected for experiment based on their final distance to the target function.

\subsection{Time-dependent Results Across All Tasks}
\label{app_ssec:AC_allresults}

Figure \ref{app_fig:ac_all_task} summarizes the experimental performance of the time-dependent QCS protocols across all six classification tasks for both amplitude and frequency-discrimination. 

Similar to the static magnetic field sensing case, for the amplitude-classification, we ran 17, 17, 24, 14, 22, and 22 protocols for tasks 1 to 6, respectively, to probe a broader range of circuit depths. For the frequency-classification we test 15, 12, 16, 16, 17, and 17 protocols for tasks 1 to 6, respectively. For our baseline, we test 6 of the best-performing dynamical decoupling sequences in simulation (See Appendix~Fig.~\ref{app_fig:supp_dd_training} for details).

The top panel shows the amplitude-classification results, while the bottom panel shows the frequency-classification results. 
In each case, the learned QSP response reproduces the target classification function with increasing fidelity as the protocol becomes longer. 
The lower rows further show that this improved matching translates into higher experimental classification accuracy as a function of total sensing time.
Across most tasks, except for the Task 1, the learned QSP protocols outperform the optimized dynamical-decoupling-based baselines over the same sensing-time range, demonstrating that task-specific learned quantum control provides a consistent advantage for time-dependent signal classification.

\subsection{Information-theoretic measure}
\label{app_ssec:information_theory}
At each step of the protocol, the quantum states are mapped onto the Bloch sphere and grouped by class. 
We  identify the measurement axis that gives the best one-shot separation between the two classes.
The optimal measurement axis is found by scanning a grid of directions over the Bloch sphere, projecting both classes onto each candidate axis, and evaluating how well the resulting one-dimensional distributions can be separated after including finite-shot sampling noise. 
Classification is then performed with a histogram-based one-dimensional classifier that bins the projected outcomes for the two classes and assigns labels by comparing the corresponding bin populations, using balanced accuracy as the figure of merit.
At each layer, we estimate the mutual information between the encoded quantum data and different target variables using the NPEET package, either from the optimal one-dimensional measurement outcome or from the full Bloch vector. 
This is done separately for the task label, the phase, and the amplitude to quantify how much relevant information is concentrated into the measurement basis and how much nuisance information remains distributed in the full state.

\section{Time-Dependent DD Protocol}
\label{app_sec:AC_DD}

\begin{figure}[htb]
    \centering    \includegraphics[width=0.65\linewidth]{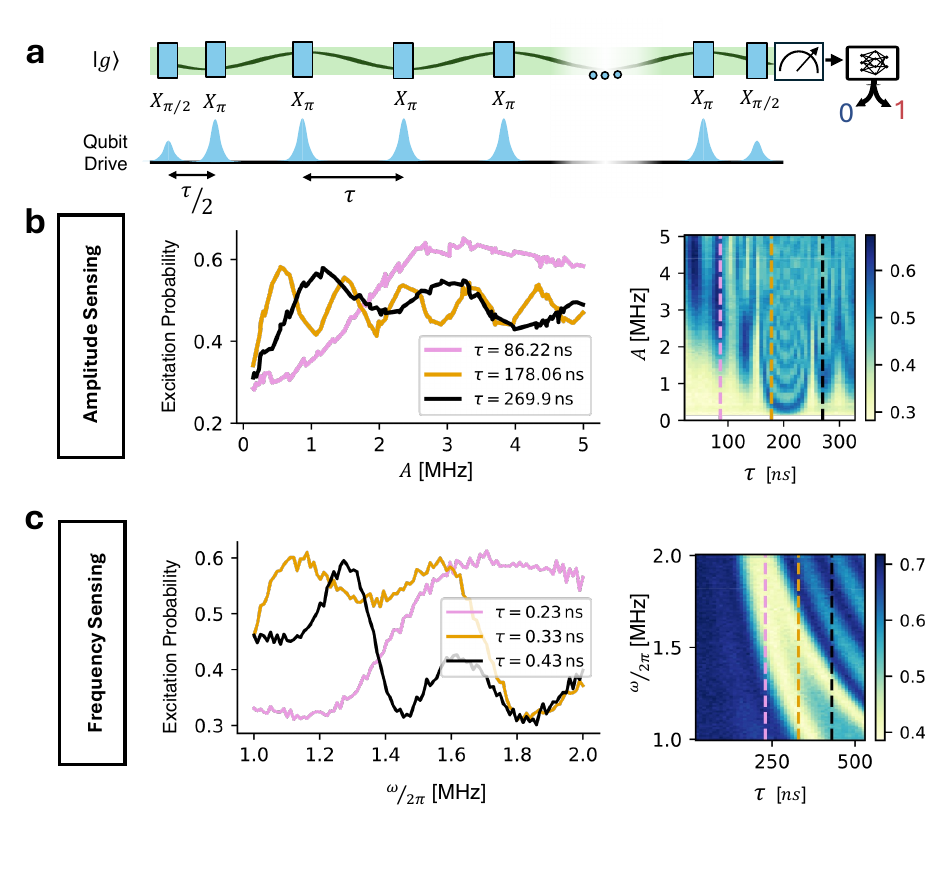}
    \caption{\textbf{Dynamical-decoupling protocols as limited function-computing sensors for time-dependent signals.} 
    \textbf{a}, Schematic of the dynamical-decoupling (DD) protocol used for time-dependent classification. 
    The sequence consists of repeated $X_\pi$ pulses separated by free-evolution intervals $\tau$. 
    By varying $\tau$, the DD response can be shaped to approximate task-dependent classification functions, showing that DD can also perform a limited form of analog computation. 
    \textbf{b}, Experimental DD response as a function of signal amplitude and wait time $\tau$ for a fixed signal frequency. 
    The red lines mark three selected values of $\tau$, labeled 1 to 3, and the corresponding one-dimensional response curves are shown in the three panels on the left. 
    These slices illustrate how different choices of $\tau$ produce different response functions that can be used for amplitude-discrimination tasks. 
    \textbf{c}, Experimental DD response as a function of signal frequency and wait time $\tau$ for a fixed signal amplitude. 
    As in b, the red lines mark three selected values of $\tau$, labeled 1 to 3, with the corresponding one-dimensional response curves shown in the three panels on the left.
    These slices illustrate the range of response functions accessible for frequency-discrimination tasks.
    }
    \label{app_fig:supp_dd_exp}
\end{figure}
\subsection{Experimental Details}

The dynamical-decoupling (DD) protocol used for time-dependent classification consists of a sequence of $\pi$-pulses separated by free-evolution intervals, as shown in Fig.~\ref{app_fig:supp_dd_exp}a. 
The protocol is parameterized by the number of pulses and the interpulse delay $\tau$, which together determine the total sensing time and the spectral response of the sequence. 
By varying these parameters, the DD protocol can generate different one-dimensional response functions of the applied signal and can therefore be optimized for a given classification task. 
Figure \ref{app_fig:supp_dd_exp}b,c shows examples of DD response functions for a fixed number of DD layers ($N_\pi=10$ for amplitude sensing and $N_\pi=5$ for frequency sensing) and different wait times $\tau$. 
These examples illustrate how varying the interpulse delay reshapes the effective response of the protocol to the applied time-dependent signal, allowing different classification functions to be approximated within the DD framework.

\begin{figure}
    \centering
    \includegraphics[width=0.75\linewidth]{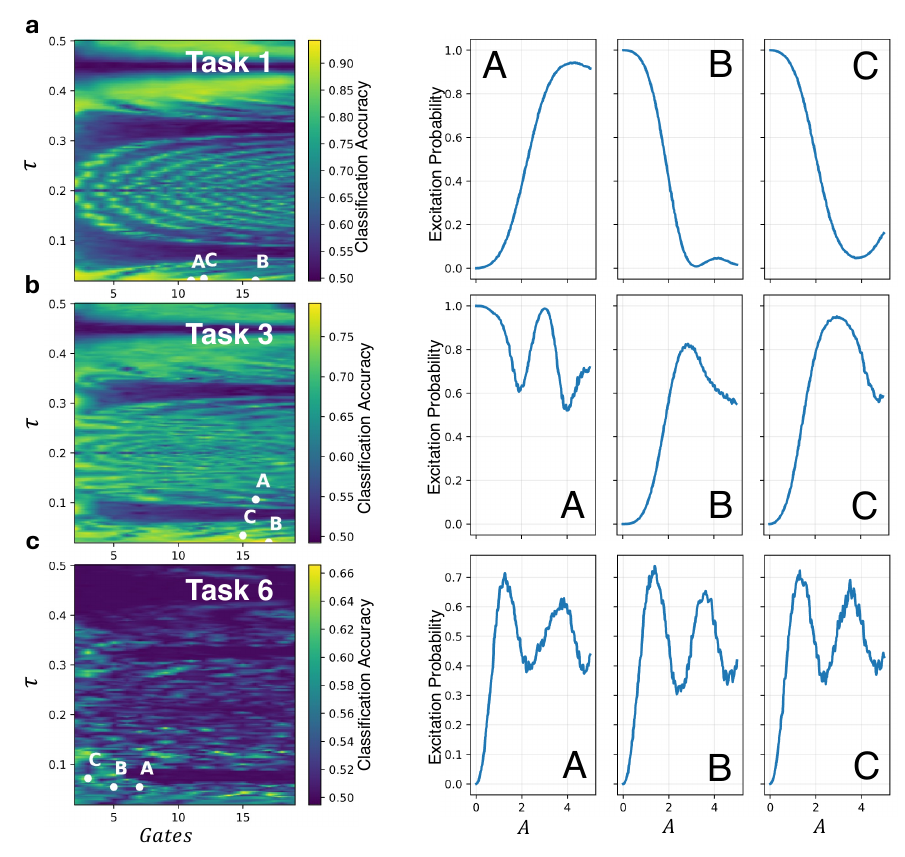}
    \caption{\textbf{Optimization of DD protocols for experimental amplitude-classification tasks.}  \textbf{a-c}, Simulated classification accuracy of dynamical-decoupling (DD) protocols as a function of pulse number and wait time for Tasks 1, 3, and 6, respectively. For each task, the optimization was carried out under a fixed total acquisition-time budget chosen to match that of the corresponding learned QSP protocol with $L=12$ and 8 shots, giving 37.75 \si{\micro\second}, 41.92 \si{\micro\second}, and 37.59 \si{\micro\second} for Tasks 1, 3, and 6, respectively. Because individual DD protocols are shorter, the number of shots for each candidate DD sequence was increased beyond 8 whenever allowed within the same total time budget. The points labeled A--C indicate representative high-performing DD protocols selected from this optimization landscape, and the corresponding response curves are shown on the right. For simple tasks, the DD optimization identifies protocols whose responses approximate the target classification structure well. For more complex tasks, however, the best accessible DD responses remain limited, resulting in lower classification accuracy and poorer agreement with the target function.}
    \label{app_fig:supp_dd_training}
\end{figure}

Unless stated otherwise, all other aspects of the experimental and numerical procedure were handled in the same way as for the QSP protocols described above. In particular, the same signal generation, calibration procedure, readout processing, shot-budget treatment, classifier training, and test-set evaluation were used, so that the DD and QSP results can be compared directly under matched conditions.

\subsection{Training Details}
\label{app_ssec:DD_training}

To identify the optimal dynamical decoupling (DD) protocol under the same total time budget, we scanned DD sequences over the number of $\pi$-pulses and the interpulse delay, and for each candidate, determined the maximum number of repetitions allowed within that budget. 
Candidates allowing fewer than one repetition were discarded. See Fig.~\ref{app_fig:supp_dd_training}.

For each scanned DD protocol, we simulated the measurement response over randomly sampled amplitudes, with projective outcomes drawn from the corresponding excited-state population. 
These outcomes were computed from the average $z$ value over a randomly sampled AC-signal phase.
Readout errors were included through a symmetric two-outcome confusion matrix with tunable fidelity. 
These measurement records were used to train and evaluate a random forest classifier, from which we computed the balanced classification accuracy under a 50:50 class prior. 
Repeating this procedure over the full DD parameter grid yielded an accuracy landscape as a function of pulse number and interpulse delay, from which the best-performing protocols were identified as shown in Fig.~\ref{app_fig:supp_dd_training}.

\subsection{Protocol structure and readout robustness.}
\label{app_ssec:DD_readout}

In this section, we study the robustness of the protocol to readout accuracy and quantify how measurement information builds up across layers.

\subsubsection{Effect of readout}

\begin{figure}
    \centering
    \includegraphics[width=\linewidth]{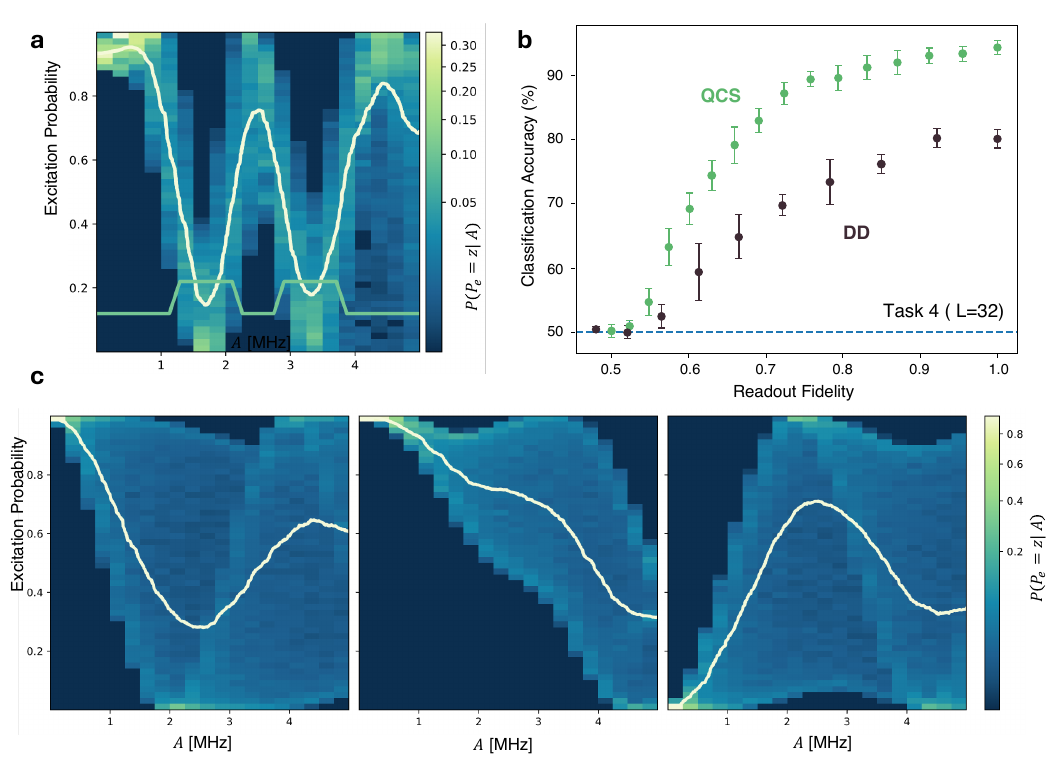}
    \caption{\textbf{Protocol structure and readout robustness.}  \textbf{a}, Optimized QCS protocol for Task 4 of oscillating-field amplitude-classification. The white curve shows the average excited-state population $P_e$ over random signal phases, and the colormap shows the corresponding phase-resolved distribution for the same protocol. The protocol largely refocuses the phase-induced spread while maintaining task-relevant dependence on amplitude. \textbf{b}, Simulated classification accuracy versus readout accuracy for QCS and DD on Task 4 ($L=32$). Although both protocols are affected by readout errors, their relative ordering remains unchanged, with QCS outperforming DD throughout. \textbf{c}, Three example DD protocols illustrating distinct strategies. The first and third aggregate the broad central response feature, while the second is closer to a Ramsey-like parameter-estimation approach.}
    \label{app_fig:supp_readout}
\end{figure}

Readout error degrades classification performance even for an ideal QCS protocol. 
In the limit where the protocol maps the two classes exactly onto different measurement eigenstates, imperfect readout still produces classification errors through outcome misassignment. 
The baseline protocol is also affected by readout error, which raises the question of whether it could outperform QCS at sufficiently low or high readout fidelity. 
Because the readout fidelity in our experiment cannot be improved beyond the optimized operating point, this question must be addressed numerically. 
To do so, we extend the shot-based simulation introduced above, which already includes phase stochasticity, by adding a symmetric confusion matrix to model readout errors. 
Figure \ref{app_fig:supp_readout} shows that, for the amplitude task 4, changes in readout fidelity do not alter the relative ordering of the two protocols. Once both methods rise above random guessing, they remain separated by a comparable margin over the full range of readout fidelities considered. This conclusion holds across all tasks considered.

Another important aspect is the refocusing ability of the different protocols, that is, their ability to suppress phase dependence and map all phases onto the same final measurement outcome. 
Figure \ref{app_fig:supp_readout} shows the distribution of $\langle z \rangle$ for different signal amplitudes, where the heat map represents the probability for the qubit to end in a given $z$ projection for both DD and QCS. 
Both protocols refocus well at zero amplitude, as expected in the absence of any amplitude-dependent effect.
This behavior, however, is rapidly lost for DD, for which the final $\langle z \rangle$ distribution extends over nearly the full range for most amplitudes.
By contrast, QCS maintains a much narrower distribution, indicating substantially stronger refocusing.
This tighter concentration directly improves class separation by reducing the intra-class spread associated with phase fluctuations.

\subsubsection{Information theory}

To better explain why QCS provides an advantage, we investigate the information-theoretic aspects of the protocol.

For both QSP and DD, the information-theory curves were obtained with exactly the same continuous-state procedure. 
Starting from the simulated wavefunction at each stage of the protocol, we converted the qubit state into Bloch-sphere coordinates $(x,y,z)$ and grouped the samples according to the task label, while keeping all amplitudes and phases used in the dataset. 
At every layer, this gives the full ensemble of qubit states produced by the protocol over the full signal distribution.

To quantify how much task-relevant information is available in a projective readout, we first identified -- independently at each layer -- the measurement axis that best separates the two classes. 
This was done by scanning candidate axes on the Bloch sphere, projecting all states onto each axis, and selecting the axis that maximized class separability. 
The resulting scalar coordinate along this optimal axis is the continuous readout variable used in the main mutual information curve. 
In parallel, we also retained the full three-dimensional Bloch vector $(x,y,z)$ to estimate how much information is present in the state before restricting to a single measurement basis.

The mutual information was then estimated layer by layer using continuous estimators. 
For the main curve, we computed the mutual information between the continuous projection onto the optimal axis and the binary target function $F^\star$, namely the class label of the task. 
For comparison, we also computed the mutual information between the full Bloch vector and the same target.

\begin{figure}[htb]
    \centering
    \includegraphics[width=0.5\linewidth]{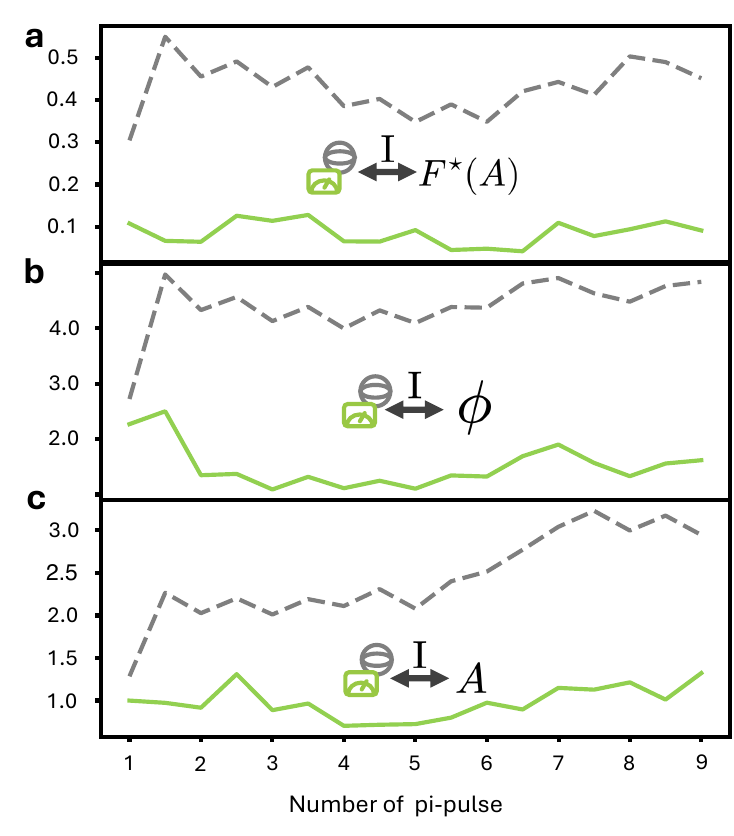}
    \caption{\textbf{Mutual information of the dynamical decoupling protocol}  \textbf{a}, Mutual information during the dynamical decoupling protocol  (middle protocol of Figure \ref{app_fig:supp_readout}c) with the target function $f(A)$ of task 4, shown after each $\pi$-pulse and at readout. The green solid curve shows the mutual information in the measured $z$ projection, while the dashed gray curve shows the mutual information in the full Bloch vector.
    \textbf{b}, Mutual information during the same protocol with the phase $\phi$.
    \textbf{c}, Mutual information during the same protocol with the amplitude $A$.
    }
    \label{app_fig:supp_info_DD}
\end{figure}

Using the same construction, we also evaluated the mutual information between the state and the signal phase $\phi$ and the signal amplitude $A$. 
For each layer, we therefore computed $I(z_{\mathrm{opt}};\phi)$ and $I(z_{\mathrm{opt}};A)$ from the continuous projection on the optimal axis, as well as $I((x,y,z);\phi)$ and $I((x,y,z);A)$ from the full Bloch vector. 
This makes it possible to track whether the protocol progressively transfers task-relevant information toward the measurement basis while suppressing sensitivity to variables that are irrelevant for the classification problem.
Figure \ref{app_fig:supp_info_DD} shows the corresponding information-theoretic curves for the DD protocols, which were omitted from the main text for readability.

\end{document}